\documentclass[letter,12pt]{article}%
\usepackage{booktabs}
\usepackage{amsmath,bm}
\usepackage{amsfonts}
\usepackage{amssymb}
\usepackage{graphicx}
\usepackage{rotating}
\usepackage{dsfont}
\usepackage[height=9in,left=1in,right=0.75in,bottom=1in]{geometry}
\usepackage[breaklinks=true,bookmarksopen=true,colorlinks=true,citecolor=blue]%
{hyperref}
\usepackage[authoryear]{natbib}
\usepackage{color}
\usepackage{colortbl}
\usepackage{paralist}
\usepackage{lscape} 
\usepackage{amsmath}%
\usepackage[formats]{listings}
\usepackage{lstautogobble}

\lstdefineformat{R}{~=\( \sim \)}
\usepackage{fancybox}

\makeatletter
\newenvironment{CenteredBox}{%
\begin{Sbox}}{
\end{Sbox}\centerline{\parbox{\wd\@Sbox}{\TheSbox}}}
\makeatother

\providecommand{\U}[1]{\protect\rule{.1in}{.1in}}
\RequirePackage{hypernat}

\catcode`~=11
\newcommand{\urltilde}{\kern -.15em\lower .7ex\hbox{~}\kern .04em}
\catcode`~=13
\def \@seccntformat#1{\csname the#1\endcsname.\quad}
\makeatother
\numberwithin{equation}{section}
\hypersetup{pdftitle={Terschuur}, pdfsubject={IOp in Europe}, pdfauthor={Joël Terschuur}, pdfkeywords={Two-step estimation; Inequality of Opportunity;
orthogonal moments, Gini, Mean Logarithmic Deviation} }
\begin{document}

\title{Educational Inequality of Opportunity and Mobility in Europe\thanks{Research funded by Ministerio de Ciencia e Innovaci\'{o}n,
grant ECO2017-86675-P, MCI/AEI/FEDER/UE, grant PGC 2018-096732-B-100, and
Comunidad de Madrid, grants EPUC3M11 (VPRICIT) and H2019/HUM-589.}}
\author{Jo\"{e}l R. Terschuur\\\textit{Universidad Carlos III de Madrid}}
\date{\today}
\maketitle

\begin{abstract}
    Educational attainment generates labor market returns, societal gains and has intrinsic value for individuals. We study Inequality of Opportunity (IOp) and intergenerational mobility in the distribution of educational attainment. We propose to use debiased IOp estimators based on the Gini coefficient and the Mean Logarithmic Deviation (MLD) which are robust to machine learning biases. We also measure the effect of each circumstance on IOp, we provide tests to compare IOp in two populations and to test joint significance of a group of circumstances. We find that circumstances explain between 38\% and 74\% of total educational inequality in European countries. Mother's education is the most important circumstance in most countries. There is high intergenerational persistence and there is evidence of an educational Great Gatsby curve. We also construct IOp aware educational Great Gatsby curves and find that high income IOp countries are also high educational IOp and less mobile countries.
    \\
    \textbf{R package:} \url{https://joelters.github.io/home/code/}
\end{abstract}

\newpage

\section{Introduction}
The acquisition of education provides labor market returns, intrinsic benefits to individuals and is valuable for society. Moreover, the distribution of educational attainment has important consequences for the distribution of economic variables. If human capital is distributed unevenly, the returns to education will feedback this inequality into economic disparities. Moreover, if the accumulation of human capital depends on characteristics out of the control of the individual, henceforth circumstances, then large parts of society are excluded from the benefits of education leading to a misallocation of human talents. Another consequence is that of inequalities in the benefits stemming from education such as health advantages (\cite{lleras2005relationship}) or life satisfaction (\cite{powdthavee2015s}) and that of reduced social gains such as less crime (\cite{lochner2020education}) and more civic participation (\cite{lochner2011non}).

In this paper we study the share of inequality in educational attainment which can be explained by circumstances, i.e. Inequality of Opportunity (IOp)\footnote{For the origins of this concept in political philosophy see \cite{rawls1971theory}, \cite{dworkina1981equality}, \cite{arneson1989equality} and \cite{cohen1989currency}. For a formalization of the IOp concept in economics see \cite{vandegaer1993}, \cite{roemer1993pragmatic}, \cite{fleurbaey1995three} or \cite{roemer1998equality}. For reviews of the empirical literature quantifying IOp, see \cite{roemer2016equality},
\cite{ramos2016approaches}, \cite{ferreira2016individual} or \cite{palmisano2022inequality}.}. We care about this component of inequality not only because it is usually considered an unfair component of inequality, but also because of its connection with intergenerational mobility and the distribution of economic variables such as income. The literature has emphasized the importance of circumstances such as parental economic resources or education, for instance \cite{coleman1968equality}, \cite{chetty2017mobility}, \cite{hallsten2018multiple} or \cite{erikson2019does}. We make use of a rich set of circumstances about parental endowments and households characteristics when the individual was $14$ years old. Applying a battery of machine learning  (ML) methods to these circumstances we are able to explain a large share of educational attainment inequality with circumstances. 

The most popular measure of IOp is the inequality in the distribution of fitted values or predictions of an outcome (e.g. education, income or health) given some circumstances. If we can get perfect predictions given circumstances, then IOp is equal to the inequality in the outcome. Hence, measuring IOp involves a two-step procedure. First, the prediction step where the outcome is forecasted and second; the fitted values inequality measurement step.

In the prediction step, ML techniques which can handle high dimensional problems, have been recently employed to make use of a rich set of circumstances without imposing any parametric assumptions. \cite{brunori2019inequality}, \cite{brunori2021roots}, \cite{brunori2021evolution}, \cite{salas2022inheritances} or \cite{carranza2022} use conditional inference forests in the prediction step and \cite{hufe2022lower} use Lasso. 

The aforementioned literature is based on an ad-hoc use of ML for the prediction stage which lacks any inferential theory for the resulting IOp measure. \cite{escanciano2022debiased} show that the use of ML techniques in the prediction stage can cause sizeable biases in the measurement of IOp. The reason is that ML trades off bias and variance, meaning that bias in the first step will be allowed if it improves the prediction. This bias in the first step might creep in the estimation of IOp in the second step. \cite{escanciano2022debiased} propose a simple debiased estimator for the Gini of the fitted values together with its inferential theory. Their results hold for general non-parametric and ML estimators which satisfy mild mean square consistency conditions.

In the fitted values inequality measurement step, the researcher needs to choose an inequality measure. Part of the literature followed \cite{checchi2010inequality} in using the Mean Logarithmic Deviation (MLD). This inequality index has the advantage of being additively decomposable (see \cite{shorrocks1980class}) into a within/between group inequality; where between inequality is precisely IOp. Other researchers prefer to use the Gini which is not additively decomposable but is more sensitive to changes in the middle of the distribution which has advantages when measuring inequality in fitted values distributions compared to the MLD which focuses on the tails of the distribution (see \cite{brunori2019inequality}). For completion we provide the form of the debiased IOp estimator based on the MLD using the theory in \cite{chernozhukov2021locally}.

The main methodological contributions of this paper are (i) to propose a measure of partial effects for individual circumstances for debiased IOp estimators, (ii) a comparison test and a group test to check whether IOp significantly differs between two independent populations and to check significance of a group of circumstances and (iii) to allow for the use of the MLD in the debiased IOp estimator. We also provide the R package \texttt{ineqopp} which allows to estimate IOp using both the Gini and the MLD and a variety of machine learners. It also computes partial effects and tests. We give examples of its usage throughout the paper.



Quantification of educational IOp can be found for educational achievement (i.e. grades) in \cite{gamboa2012inequality} who study IOp in 2006-2009 PISA scores in Latin America or in \cite{lasso2020reexamining} who measure IOp in PISA 2012 scores in 20 European countries. IOp in access to higher education can be found in \cite{krafft2018inequality} for the Middle East and North Africa or in \cite{palmisano2022inequality} who study IOp in access to tertiary education in Europe using the EU-SILC waves of 2005 and 2011. In this paper we focus in years of education and do not consider other qualitative differences such as the quality of the educational institutions attended or the field of study. We do not study these qualitative differences because of lack of data and not because they are not important (see \cite{BLANDEN2023405}).



We use the 2019 cross-sectional European Union Statistics on Income and Living Conditions (EU-SILC) survey to study IOp in years of education in European countries. We employ regularized linear regression techniques such as Lasso and Ridge and tree-based ensemble methods such as Random Forests (RF), Conditional Inference Forest (CIF), Extreme Gradient Boosting (XGB) and Catboosting (CB)\footnote{Lasso, Ridge and RF are well-known MLs in econometrics. CIF were developed as an alternative to RF in
\cite{hothorn2006unbiased} and they are used by the IOp literature. XGB and CB are boosting algorithms which have become popular in machine learning competitions (see \cite{chen2016xgboost} and \cite{prokhorenkova2018catboost}).} to exploit a rich set of circumstances. We find that the share of educational inequality explained by circumstances ranges from 38\% in Finland to 74\% in Bulgaria and that plug in estimates (ad-hoc use of machine learners) are about 3 to 10 times more sensitive to first step machine learners compared to the debiased counterparts. These estimates are large compared to similar estimates in the literature. \cite{erikson2019does} explains no more than 25\% of the variation in educational attainment with social origin. When we divide the sample into a young and an old cohort we do not see much change in the range of the estimates although we see that IOp has increased significantly in the young cohort in 10 countries out of the 30 countries. 

When we look at the impact of different circumstances the main result that we get is that mother's education plays a vital role in predicting educational attainment. In 17 out of the 30 countries mother's education has the highest effect. For instance, in Sweden, excluding mother's education decreases IOp a 13\%. The next most important circumstance seems to be sex which is the most important circumstance in Nordic and Baltic countries. This seems to be driven by the superior educational attainment of females compared to males in these countries.

Motivated by the importance of mother's education, we turn to study intergenerational mobility. We analyse how maternal educational attainment persists in the educational attainment of the individual. We find evidence for low educational mobility with similar patterns as those for educational IOp. Following \cite{blanden2013cross}, \cite{corak2013income}, \cite{oecd2018} and \cite{BLANDEN2023405}, we look at the association between mobility and income inequality. As in these studies we find evidence for an educational Great Gatsby curve, i.e. countries with high income inequality tend to have less educational mobility. Next, we substitute income inequality in the Great Gatsby curve by income IOp and we see that the relationship becomes tighter and the slope becomes more positive. This indicates that circumstances matter in the relationship between educational mobility and income inequality. We would expect this to happen if highly educated mothers do not only affect child's education through monetary investments and their own education but also pass on some latent status as studied in \cite{stuhler2023}. If this latent status is correlated with the circumstances we observe then we would expect this relationship between mobility and income IOp. 

We also look at the correlation between educational IOp and income IOp and find a strong, positive association, i.e. we see no countries with high income IOp and low educational IOp or vice versa. Instead, high income IOp countries tend to have high educational IOp. This means that reducing educational IOp without reducing income IOp can be particularly challenging.

Next, we look at IOp differences between the population of males and females. We argue this is a good characterization of an IOp gender gap since biological sex is randomly assigned and circumstances are predetermined. This means that the distribution of circumstances is the same for females and males so the difference in IOp cannot be attributed to composition effects but to differences in returns to circumstances. This could happen for instance if parental education and occupation have different impacts on girls and boys due to gendered role models. While females are obtaining more educational attainment than males on average, we do not find any clear patterns when we compare educational IOp among females and males. This is in line with previous findings in \cite{erikson2019does}.

Finally, we explore some further associations with other variables such as average years of education, average income, educational inequality and education expenditure as a percentage of GDP. We find that educational IOp decreases with per capita educational attainment, average income and education expenditure and increases with educational inequality. However, the last correlation is weak and we still find countries with high educational inequality and low educational IOp and vice versa relative to other countries.

Section \ref{sec_deb_iop} introduces the Gini and MLD debiased estimators, Section \ref{sec_pe} defines and provides the partial effects and their estimators, Section \ref{sec_tests} provides the comparison and group test, Section \ref{sec_IOp_Europe} shows the empirical application to the EUSILC survey, Section \ref{sec_discussion} discusses limitations and Section \ref{sec_conclusion} concludes.

\section{Debiased IOp}
\label{sec_deb_iop}
Let $W_i = (Y_i,X_i)$ be i.i.d. data distributed as $F_0$. $Y_i$ is a positive scalar-valued outcome such as years of education and $X_i$ is a vector of circumstances. Let $\mathbb{E}_F$ be the expectation under distribution $F$, $\mathbb{E} \equiv \mathbb{E}_{F_0}$ and denote $\gamma_0(X_i) = \mathbb{E}_{F_0}[Y_i |X_i]$ as the true fitted values under $F_0$. We are interested in the inequality of the distribution of the random variable $\gamma_0(X_i)$. Two available measures to quantify this inequality are the Gini and the MLD.

\subsection{Debiased Gini IOp}
\label{sec_gini_iop}
The Gini of the fitted values is
\begin{equation}
\theta_0 (\gamma_0) = \frac{\mathbb{E}[|\gamma_0(X_i) - \gamma_0(X_j)|]}{\mathbb{E}[\gamma_0(X_i) + \gamma_0(X_j)]}.
\label{eq_gini_FVs}
\end{equation}
Estimation of $\theta_0$ can be framed as a method of moments problem. From (\ref{eq_gini_FVs}) we get
\begin{equation}
\mathbb{E}[\theta(\gamma_0(X_i) + \gamma_0(X_j)) - |\gamma_0(X_i) - \gamma_0(X_j)|] = 0, \text{ iff } \theta = \theta_0(\gamma_0).
\label{eq_moment}  
\end{equation}
The standard estimation method is the so called plug in estimator where we plug in an estimator $\hat{\gamma}$ of $\gamma_0$ in a sample version of (\ref{eq_moment}) and solve for $\theta$. For example, using Conditional Inference forests, the plug in estimate can be computed in R as

\begin{figure}[thp]
\begin{CenteredBox}
\lstset{basicstyle=\footnotesize\ttfamily,
format = R}
\begin{lstlisting}
model <- cforest(Y ~ ., data)
fitted_values <- predict(model, X)
IOp <- Gini(fitted_values)
\end{lstlisting}
\end{CenteredBox}
\end{figure}
\noindent where \texttt{data} is a data frame containing outcome $Y$ and all circumstances $X$. Using the package \texttt{ineqopp}, which we provide, this can be done as
\begin{figure}[thp]
    \begin{CenteredBox}
        \begin{lstlisting}
IOp <- IOp(Y, X, est_method = "Plugin", ineq = "Gini",
    plugin_method = "ML", ML = "CIF")
        \end{lstlisting}
    \end{CenteredBox}
\end{figure}

As indicated by the notation $\theta_0(\gamma_0)$, IOp depends on the fitted values $\gamma_0$. The problem with the plug in estimator is that it is not locally robust in the sense that
\[
\frac{\partial \theta_0(\gamma_0)}{\partial \gamma} \neq 0.
\]
Whenever the above derivative is different from zero, regularization and model selection biases from the first step will translate into biases and invalid inference of the IOp estimator. The solution to the lack of local robustness is to find an orthogonal moment condition. This is an alternative moment condition which still identifies $\theta_0$ but has zero derivative with respect to first steps. \cite{escanciano2022debiased} find the following orthogonal moment condition
\[
\mathbb{E}[\theta(Y_i + Y_j) - sgn(\gamma_0(X_i) - \gamma_0(X_j))(Y_i - Y_j)] = 0, \text{ iff } \theta = \theta_0(\gamma_0).
\]
Based on this moment condition, $\partial \theta_0(\gamma_0)/\partial \gamma = 0$, meaning that we achieve local robustness, reduce regularization and model selection biases and we are able to find a limiting asymptotic distribution on which to base inference.
The sample version is 
\begin{equation}
\sum_{i} \sum_j \theta(Y_i + Y_j) - sgn(\hat{\gamma}(X_i) - \hat{\gamma}(X_j))(Y_i - Y_j) = 0.
\label{eq_sample_orth_cond}
\end{equation}
Solving the above for $\theta$ gives

\begin{equation}
\theta = \frac{\sum_{i} \sum_j sgn(\hat{\gamma}(X_{i})-\hat{\gamma}(X_{j}))(Y_{i}-Y_{j}%
)}{\sum_i \sum_j (Y_i + Y_j)}.\label{DG}%
\end{equation}
(\ref{DG}) is simple to implement: it is like the standard Gini coefficient but with the sign of outcome differences replaced by the sign of fitted values differences.\footnote{The standard Gini coefficient is $\sum_{i} \sum_j |Y_i - Y_j|/\sum_{i} \sum_j(Y_i + Y_j) = \sum_{i} \sum_j sgn(Y_i - Y_j)(Y_i - Y_j)/\sum_{i} \sum_j(Y_i + Y_j)$.} This means that whenever two individuals have
the same fitted values, their difference in outcomes cannot be attributed to
IOp. When individual $i$'s fitted values are strictly greater than those of individual $j$, the
difference in outcomes $Y_i - Y_j$ enters positively. On the contrary, if $\hat{\gamma}(X_{i})<\hat{\gamma}(X_{j})$, then the difference in outcomes $Y_i - Y_j$ enters negatively. Hence, the two driving forces of IOp are (i) the differences in outcomes and (ii) whether these differences coincide in sign with the differences in predicted outcomes.

In (\ref{DG}) we are summing across pairs $(i,j)$ and we are using the same observations to estimate the fitted values. This can lead to upward biases due to overfitting (whenever the fitted values capture sampling noise) and complicates the theoretical inference results when using ML, see \cite{chernozhukov2018double}. Cross-fitting uses different observations to compute the sum in (\ref{DG}) and the fitted values. To perform cross-fitting in the case of the Gini we divide the pairs $(i,j)$ in blocks $I_1,...,I_L$ and let $\hat{\gamma}_l$ be the fitted values estimated with observations not in $I_l$, then the debiased estimator is  
\[
\hat{\theta} = \frac{\sum_{l=1}^L \sum_{(i,j) \in I_l} sgn(\hat{\gamma}_l(X_{i})-\hat{\gamma}_l(X_{j}))(Y_{i}-Y_{j}
)}{\sum_i \sum_j (Y_i + Y_j)}.
\]
For an example on how to create blocks $I_1,...,I_L$ see the supplementary material. \cite{escanciano2022debiased} show that
\[
\sqrt{n}(\hat{\theta} - \theta_0) \to_d \mathcal{N}(0,V),
\]
where
\[
V = \mathbb{E}[Y_i]^{-2}\mathbb{E}\biggl(\mathbb{E}[(\theta_0(Y_i + Y_j) - sgn(\gamma_0(X_i) - \gamma_0(X_j))(Y_i - Y_j))|Y_i,X_i]^2\biggr),
\]
and a consistent estimator is
\[
\hat{V}=\frac{\frac{1}{n(n-1)^{2}}\sum_{i=1}^{n}\biggl(\sum_{j\neq i}%
\hat{\theta}(Y_{i}+Y_{j})-sgn(\hat{\gamma}(X_i) - \hat{\gamma}(X_j))(Y_{i}%
-Y_{j})\biggr)^{2}}{\Bar{Y}^{2}}.
\]
Note that there is no need to use cross-fitting to estimate the variance. The standard errors can be estimated as $\hat{se} = \sqrt{\hat{V}/n}$ and $95 \%$ confidence intervals can be constructed as usual: $CI_{95} = [\hat{\theta} - z_{0.025}\hat{se}, \hat{\theta} + z_{0.975}\hat{se}]$, where $z_\alpha$ is the $\alpha$ quantile of a standard normal distribution. With the package \texttt{ineqopp} the debiased IOp estimator with its standard errors using Random Forests to estimate the fitted values can be computed as

\begin{figure}[thp]
    \begin{CenteredBox}
        \begin{lstlisting}
IOp <- IOp(Y, X, est_method = "Debiased",
    ineq = "Gini", ML = "RF")
        \end{lstlisting}
    \end{CenteredBox}
\end{figure}

\subsection{Debiased MLD IOp}
Let $\theta_{01} = \mathbb{E}[\gamma_0(X_i)]$ and $\theta_{02} = \mathbb{E}[\ln \gamma_0(X_i)]$, then the MLD of the fitted values is
\[
\theta_0 = \ln \theta_{01} - \theta_{02},
\]
i.e. the MLD is the difference between the log of the expectation and the expectation of the log. The identifying moment conditions for $(\theta_{01},\theta_{02})$ are $\mathbb{E}[\gamma_0(X_i) - \theta_{01}] = 0$ and $\mathbb{E}[\ln \gamma_0(X_i) - \theta_{02}] = 0$ respectively. From \cite{newey1994asymptotic} and \cite{chernozhukov2021locally}, the orthogonal moment conditions for $(\theta_{01},\theta_{02})$ are
\[
\mathbb{E}[Y_i - \theta_{01}] = 0, \quad \mathbb{E}\biggl[\ln \gamma_0(X_i) - \theta_{02} + \frac{1}{\gamma_0(X_i)}(Y_i - \gamma_0(X_i))\biggr] = 0.
\]
All mathematical derivations are in the Supplementary Material. Hence, the debiased estimators are 
\[
\hat{\theta}_{1} = \Bar{Y}, \quad \hat{\theta}_{2} = \frac{1}{n}\sum_{l = 1}^L \sum_{i \in I_l} \ln \hat{\gamma}_l(X_i) + \frac{1}{\hat{\gamma}_l(X_i)}(Y_i -\hat{\gamma}_l(X_i)), 
\]
where now $I_1,...,I_L$ is a partition of $\{1,...,n\}$ and the debiased MLD IOp estimator is
\[
\hat{\theta} = \ln \hat{\theta}_1 - \hat{\theta}_2.
\]
The asymptotic properties of $\hat{\theta}$ follow easily from those of $(\hat{\theta}_1,\hat{\theta}_2)$ and \cite{chernozhukov2021locally}. The asymptotic variance can be estimated as
\[
\hat{V}  = \frac{\widehat{Var}[Y_i]}{\hat{\theta}^2_1} +  
\widehat{Var}[\psi(W_i,\hat{\gamma}_l,\hat{\theta}_2)] - \frac{2\widehat{Cov}[Y_i,\psi(W_i,\hat{\gamma}_l,\hat{\theta}_2)]}{\hat{\theta}_1},
\]
where $\psi(W_i,\hat{\gamma}_l,\hat{\theta}_2) = \ln \hat{\gamma}_l(X_i) - \hat{\theta}_2 + (1/\hat{\gamma}_l(X_i))(Y_i - \hat{\gamma}_l(X_i))$. As before, the standard errors can be estimated as $\hat{se} = \sqrt{\hat{V}/n}$ and $95 \%$ confidence intervals can be constructed as usual: $CI_{95} = [\hat{\theta} - z_{0.025}\hat{se}, \hat{\theta} + z_{0.975}\hat{se}]$. With the package \texttt{ineqopp} the debiased IOp estimator with its standard errors using Lasso to estimate the fitted values can be computed as

\begin{figure}[thp]
    \begin{CenteredBox}
        \begin{lstlisting}
IOp <- IOp(Y, X, est_method = "Debiased",
    ineq = "MLD", ML = "Lasso")
        \end{lstlisting}
    \end{CenteredBox}
\end{figure}

\subsection{Relative IOp}
\label{subsec_relative_iop}
For the sake of comparability between estimates of different countries, regions or groups; we also report the following relative measure of IOp: $\theta_0^R = \theta_0/I$, where $I$ stands for some inequality index of the distribution of $Y_i$, in our case either the Gini or the MLD. $\theta_0$ refers to the IOp parameter. Whether we are employing the Gini or the MLD will be clear from the context. A consistent estimator is $\hat{\theta}^R = \hat{\theta}/\hat{I}$,
where $\hat{I}$ is the estimator of the inequality index of $Y_i$. The inference for this relative measure follows easily from the asymptotic properties of $\hat{\theta}$ and the estimator of unconditional inequality (i.e. Gini or MLD). For brevity we omit the expression of the estimators of the asymptotic variances but they can be found in equations (3.1) and (3.2) in the Supplementary Material. Relative IOp and its standard errors can be computed with the package \texttt{ineqopp} by setting the option \texttt{\text{IOp\_rel}} to \texttt{TRUE} in the \texttt{IOp} function. For instance, for the Gini and using Extreme Gradient Boosting

\begin{figure}[thp]
    \begin{CenteredBox}
        \begin{lstlisting}
IOp <- IOp(Y, X, est_method = "Debiased",
    ineq = "Gini", ML = "XGB", IOp_rel = TRUE)
        \end{lstlisting}
    \end{CenteredBox}
\end{figure}

\section{Partial effects}
\label{sec_pe}
While MLs allow to use many circumstances in the estimation of IOp, it is not entirely clear how to measure the impact that each circumstance has on IOp. \cite{brunori2021roots} uses variable importance in the CIF of the first step. This is a measure of how the CIF fit worsens whenever we exclude a variable from the set circumstances. This approach does not take into account the effect that dropping a variable has on IOp, only on the prediction of the outcome. To overcome this issue we propose to look at the difference between IOp estimates whenever we exclude a given circumstance.

Suppose that we have $M$ circumstances, i.e. $X_i \in \mathbb{R}^M$. Let $\theta_{0,-m}$, for $m = 1,...,M$, be the true IOp whenever we do not include circumstance $m$ and $\hat{\theta}_{-m}$ be the debiased IOp estimator without including circumstance $m$. Then we define the partial effect of circumstance $m$ to be
\[
\kappa_{0,m} =\theta_0 -  \theta_{0,-m}.
\]
It is known in the literature that $\kappa_{0,m} \geq 0$. That is, including more circumstances will increase IOp unless the added circumstances do not help to predict the outcome, in which case IOp will not change. This leads to the interpretation of the inequality of the fitted values being a lower bound of IOp when we do not observe all relevant circumstances. However, this lower bound property might not be true in the sample, since we also capture inequality stemming from sample noise. The partial effect for $m = 1,...,R$ can be estimated by \[
\hat{\kappa}_m = \hat{\theta} - \hat{\theta}_{-m}.
\]
The computation and estimation of standard errors and confidence intervals for the partial effects follow easily from the asymptotic properties of $(\hat{\theta},\hat{\theta}_{-m})$. The expressions for the estimators of the asymptotic variances can be found in equations (4.1) and (4.3) in the Supplementary material. This partial effects can be computed with the \texttt{ineqopp} package. For example, for the Gini and Catboosting, the partial effects from maternal and paternal education can be computed as

\begin{figure}[thp]
    \begin{CenteredBox}
        \begin{lstlisting}
circs <- c("educM","educF")
iop <- IOp(Y, X, est_method="Debiased", 
    ineq = "Gini", ML="CB", fitted_values = TRUE)
FVs <- iop$FVs
iop_gini <- iop$IOp["IOp"]
pe <- peffect(X, Y, circs, FVs = FVs, ineq = "Gini",
    ML = "CB", iop_gini = iop_gini)
        \end{lstlisting}
    \end{CenteredBox}
\end{figure}

\subsection{Relative Partial Effects}
Again, in order to better compare the magnitude of partial effects among different countries we also report the following relative measure of the impact of circumstance $m = 1,...,M$,
\[
\kappa_{0,m}^R = \frac{\theta_0 - \theta_{0,-m}}{\theta_0},
\]
which can be consistently estimated by
\[
\hat{\kappa}_m^R = \frac{\hat{\theta} - \hat{\theta}_{-m}}{\hat{\theta}}.
\]
$\kappa_{0,m}^R$ measures the decrease in IOp if we exclude circumstance $m$. For example, if $\kappa_{0,m}^R = 0.1$, it means that if we exclude circumstance $m$, then IOp decreases a $10 \%$. Expressions for the estimators of the asymptotic variances are in equations (4.2) and (4.4) in the Supplementary material. To compute relative partial effects with the \texttt{ineqopp} package set the option \texttt{\text{pe\_rel}} to \texttt{TRUE} in the \texttt{peffect} function.

\section{Tests}
\label{sec_tests}
A crucial advantage of being able to perform inference is the possibility to do statistical tests. Suppose that we have independent populations $A$ and $B$, e.g. two countries, and debiased IOp estimates $\hat{\theta}_A$ and $\hat{\theta}_B$ of true IOp $\theta_A$ and $\theta_B$ with estimated standard errors $\widehat{se}(\hat{\theta}_A)$ and $\widehat{se}(\hat{\theta}_A)$. Then, we can estimate the standard error of the difference in IOp $\hat{\theta}
_A - \hat{\theta}_B$ as
\[
\widehat{se}(\hat{\theta}_A - \hat{\theta}_B) = \sqrt{\widehat{se}^2(\hat{\theta}_A) + \widehat{se}^2(\hat{\theta}_B)}.
\]
Hence, we can test whether debiased IOp in population A is significantly different from that in population B. The function \texttt{IOptest} in the \texttt{ineqopp} package performs this test.

Another possibility is to test for significance of a group of circumstances. In fact, checking whether $0$ is included in the confidence interval for $\hat{\kappa}_m$ already constitutes a significance test for circumstance $m$. To do a group test we simply compare the debiased IOp estimate $\hat{\theta}$ with an IOp estimate $\hat{\theta}_{-[U]}$ where $U \subseteq \{1,...,M\}$ indicates which circumstances we want to test and $\hat{\theta}_{-[U]}$ is the debiased IOp estimate excluding those circumstances. In the case in which $U = \{1,...,M\}$ we are just checking whether the debiased IOp estimate is significant. The group test can be done with \texttt{ineqopp} package by using the function \texttt{IOpgrouptest}.

\section{Educational Inequality of Opportunity in Europe}
\label{sec_IOp_Europe}
We use the 2019 cross-sectional European Union Statistics on Income and Living Conditions (EU-SILC) survey to study IOp in years of education. In the years 2005, 2011 and
2019, EU-SILC includes a module on intergenerational transmission of disadvantages
with information on circumstances. We restrict our attention to the year 2019
since it is the most recent one and contains the richest set of circumstances. We also show all of our results for the Gini index, using the MLD gives lower estimates but the correlation with the Gini estimates is close to one (see Figure \ref{fig_sct_iopgini_iopmld}). All the results we show here can be found for the MLD in the Supplementary material.

We restrict the sample to those aged between 25 and 59 years old. The circumstances include questions
on characteristics of the parents and questions related to the individual's
life/household when he/she was around 14 years old. We use the following circumstances:
sex, country of birth, whether he/she was living with the mother/father, the number of adults/working adults/kids in the household, population of the municipality, tenancy of the
house, country of birth of the parents, nationality of the parents,
education of the parents, occupational status of the parents, father's managerial position, father's occupation, basic
school needs (whether he/she had access to books, materials, etc.), financial
situation, food needs (whether he/she could eat meat/chicken/fish/vegetarian
equivalent once a week and holidays outside of home once a year). Remember that
it refers to when the individual was around 14 years old, so, for instance,
financial situation refers to the financial situation of the household where
the individual resided when he/she was around 14 years old.

EU-SILC does not directly ask for years of education but it asks for the highest ISCED educational level achieved. Following \cite{checchi2014educational}, we compute years of education by setting ISCED levels $0$ and $1$ to $7$ years of education, level $2$ to $10$ years, level $3$ to $13$ years, level $4$ to $15$ years and levels $5$ or higher to $18$ years of education. 

All circumstances are categorical and there are many different combinations of
the categories. This makes the problem hard to deal with
without machine learning procedures. In this application we use Lasso,
Ridge, RF, CIF, XGB and CB. For Lasso and Ridge we use dummy encoding and employ a dictionary including up to 10-wise interactions ($2,666$ regressors) depending on each country. For cross-fitting we need to choose how many splits $K$ we do to the sample. We set $K = 5$. We use the appropriate individual cross-sectional
weights. 

\begin{figure}
    \centering\includegraphics[scale = 0.35]{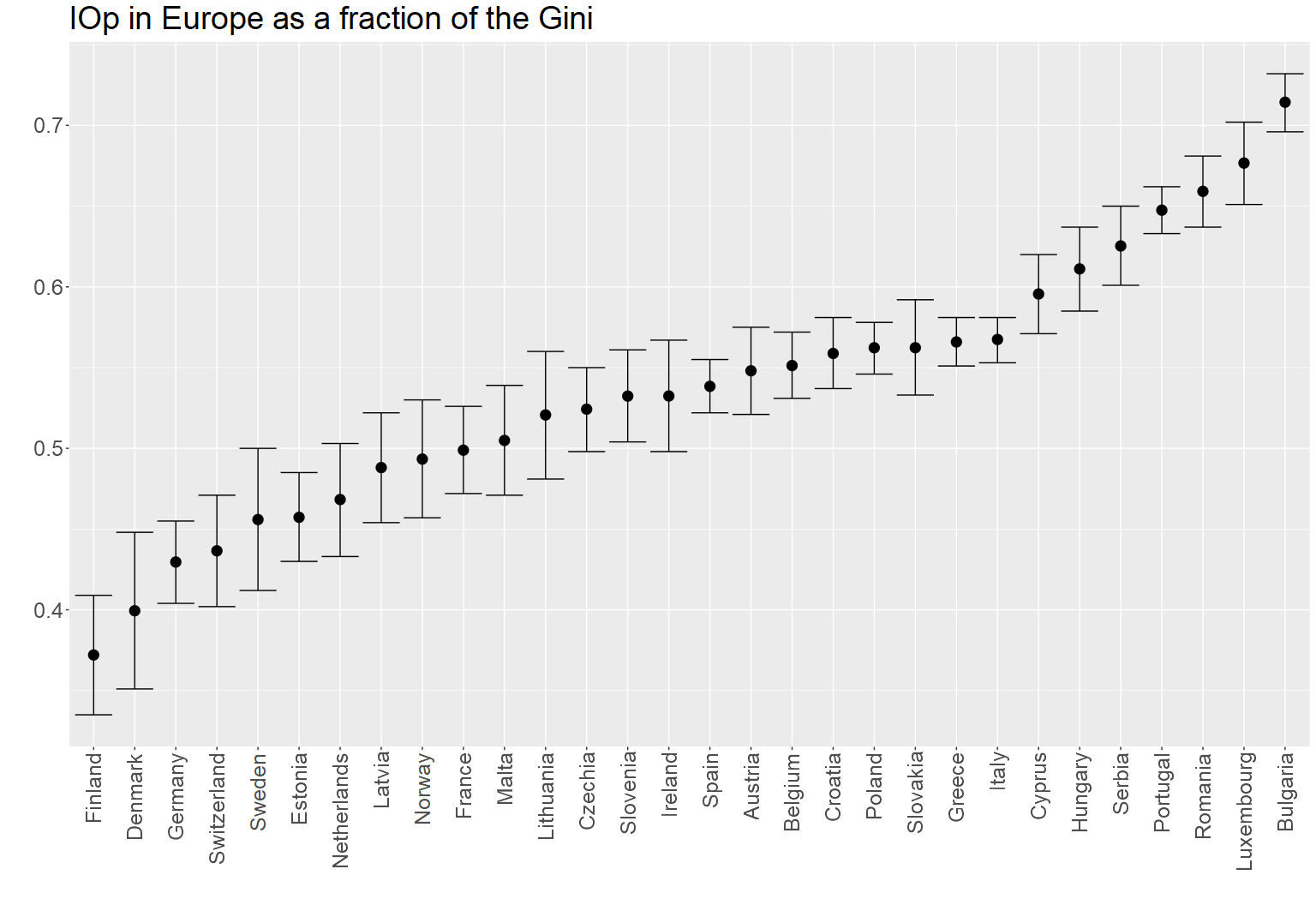}
    \caption{\small Educational IOp in Europe}
    \label{fig_educ_IOp}
\end{figure}

In Figure \ref{fig_educ_IOp} we see the first main empirical result: IOp in years of education as a share of total inequality in years of education. We estimated IOp with the 5 mentioned MLs but we report only the estimates from the MLs which achieved the lowest first stage RMSE in each country (henceforth the best ML). Hence, each country's IOp is estimated using different MLs. Relative IOp ranges from around 40\% in northern countries such as Denmark or Finland to around 70\% in countries like Romania, Portugal, Luxembourg and Bulgaria. Southern countries (with the exception of Portugal) tend to align in the middle of the ranking with almost 60\% of inequality explained by circumstances. Germany, Netherlands and France all have between 40\% and 50\% of educational IOp. Baltic countries have around 50\% of their years of education inequality explained by circumstances. Central and Eastern countries are more spread out. In the middle of the ranking we find Czechia or Poland with around 55\% of educational IOp. Countries like Slovakia or Hungary have almost 60\% of total inequality explained by circumstances while Romania and Bulgaria have around 70\%. In Table \ref{tab_educ_IOp} we report detailed results.

\begin{landscape}
\begin{table}[h!]
\footnotesize
\begin{tabular}{@{}ccccccccc@{}}
\toprule
Country & Mean & Gini & Debiased & Debiased/Gini & Estimates std. ratio & Best Performer & RMSE & n \\ \midrule
Austria & 14.6 & 0.103 & 0.057 (0.053,0.06) & 55 \%  (52\%,57\%) & 7.7 & Lasso & 2.48 & 5121 \\
Belgium & 14.8 & 0.123 & 0.068 (0.064,0.071) & 55 \%  (53\%,57\%) & 17.1 & CIF & 2.95 & 6383 \\
Bulgaria & 13.9 & 0.115 & 0.082 (0.079,0.085) & 71 \%  (70\%,73\%) & 4.9 & CIF & 2.29 & 5906 \\
Switzerland & 15.3 & 0.095 & 0.042 (0.038,0.045) & 44 \%  (40\%,47\%) & 11.3 & XGB & 2.61 & 4709 \\
Cyprus & 14.5 & 0.126 & 0.075 (0.071,0.079) & 60 \%  (57\%,62\%) & 15.3 & CIF & 2.88 & 4647 \\
Czechia & 14.1 & 0.077 & 0.04 (0.038,0.043) & 52 \%  (50\%,55\%) & 3.7 & XGB & 2.11 & 6883 \\
Germany & 15 & 0.095 & 0.041 (0.038,0.044) & 43 \%  (40\%,46\%) & 17.1 & CIF & 2.42 & 5985 \\
Denmark & 15.2 & 0.103 & 0.041 (0.036,0.047) & 40 \%  (35\%,45\%) & 6.4 & XGB & 2.82 & 2079 \\
Estonia & 15.1 & 0.101 & 0.046 (0.043,0.049) & 46 \%  (43\%,48\%) & 6.2 & CIF & 2.59 & 5666 \\
Greece & 13.9 & 0.142 & 0.08 (0.077,0.083) & 57 \%  (55\%,58\%) & 7.5 & CIF & 3.04 & 14227 \\
Spain & 13.7 & 0.157 & 0.085 (0.082,0.088) & 54 \%  (52\%,56\%) & 7 & CIF & 3.36 & 16892 \\
Finland & 15.1 & 0.096 & 0.036 (0.032,0.04) & 37 \%  (34\%,41\%) & 7.3 & Lasso & 2.63 & 4399 \\
France & 14.7 & 0.118 & 0.059 (0.055,0.063) & 50 \%  (47\%,53\%) & 7.2 & XGB & 2.94 & 7877 \\
Croatia & 13.8 & 0.092 & 0.052 (0.049,0.054) & 56 \%  (54\%,58\%) & 7.6 & CIF & 2.28 & 7126 \\
Hungary & 14.1 & 0.101 & 0.062 (0.058,0.066) & 61 \%  (58\%,64\%) & 6.9 & CIF & 2.25 & 4563 \\
Ireland & 15.6 & 0.107 & 0.057 (0.052,0.062) & 53 \%  (50\%,57\%) & 7.6 & CIF & 2.95 & 3653 \\
Italy & 13 & 0.125 & 0.071 (0.068,0.073) & 57 \%  (55\%,58\%) & 9.8 & CIF & 2.58 & 16380 \\
Lithuania & 15.3 & 0.091 & 0.047 (0.043,0.051) & 52 \%  (48\%,56\%) & 7.9 & XGB & 2.27 & 3648 \\
Luxembourg & 14.6 & 0.132 & 0.089 (0.084,0.095) & 68 \%  (65\%,70\%) & 4 & CIF & 2.8 & 3521 \\
Latvia & 14.9 & 0.099 & 0.049 (0.045,0.052) & 49 \%  (45\%,52\%) & 3.8 & XGB & 2.47 & 3351 \\
Malta & 13.1 & 0.138 & 0.07 (0.065,0.075) & 50 \%  (47\%,54\%) & 10.6 & XGB & 2.97 & 3625 \\
Netherlands & 15.3 & 0.102 & 0.048 (0.044,0.052) & 47 \%  (43\%,50\%) & 5.3 & XGB & 2.8 & 4462 \\
Norway & 14.3 & 0.141 & 0.07 (0.063,0.076) & 49 \%  (46\%,53\%) & 8.1 & Ridge & 3.33 & 2367 \\
Poland & 14.4 & 0.104 & 0.058 (0.056,0.061) & 56 \%  (55\%,58\%) & 7.2 & CIF & 2.59 & 14004 \\
Portugal & 12.1 & 0.19 & 0.123 (0.12,0.126) & 65 \%  (63\%,66\%) & 9.2 & CB & 3.22 & 13779 \\
Romania & 13.4 & 0.102 & 0.067 (0.064,0.071) & 66 \%  (64\%,68\%) & 8.1 & CIF & 2.17 & 5961 \\
Serbia & 13.8 & 0.101 & 0.063 (0.059,0.066) & 63 \%  (60\%,65\%) & 4.9 & XGB & 2.32 & 5608 \\
Sweden & 14.9 & 0.11 & 0.05 (0.044,0.056) & 46 \%  (41\%,50\%) & 5.7 & CIF & 2.79 & 1897 \\
Slovenia & 14.6 & 0.098 & 0.052 (0.049,0.056) & 53 \%  (50\%,56\%) & 2.8 & CIF & 2.43 & 4432 \\
Slovakia & 14.1 & 0.08 & 0.045 (0.042,0.048) & 56 \%  (53\%,59\%) & 4.2 & XGB & 2.09 & 5731 \\ \bottomrule
\end{tabular}
\caption{\small Educational Gini IOp: debiased vs. plug in estimators, ratio of the standard deviation of plug in estimates and the standard deviation of debiased estimates across different MLs, and performance of MLs.}
\label{tab_educ_IOp}
\end{table}
\end{landscape}

As expected from the robustness of debiased estimators, plug in estimators are much more sensible to the choice of the machine learner. The standard deviation of plug in estimates across MLs is from 2.8 times to 10.6 times the standard deviation of debiased estimates across MLs, i.e. the standard deviation is taken across the estimates resulting from different MLs. Finally, CIF was the best ML in 15 countries followed by XGB which was the best in 10 countries. Lasso was the best in two countries and Ridge and CB were best in one country.

\begin{figure}[!htb]
    \centering\includegraphics[scale = 0.3]{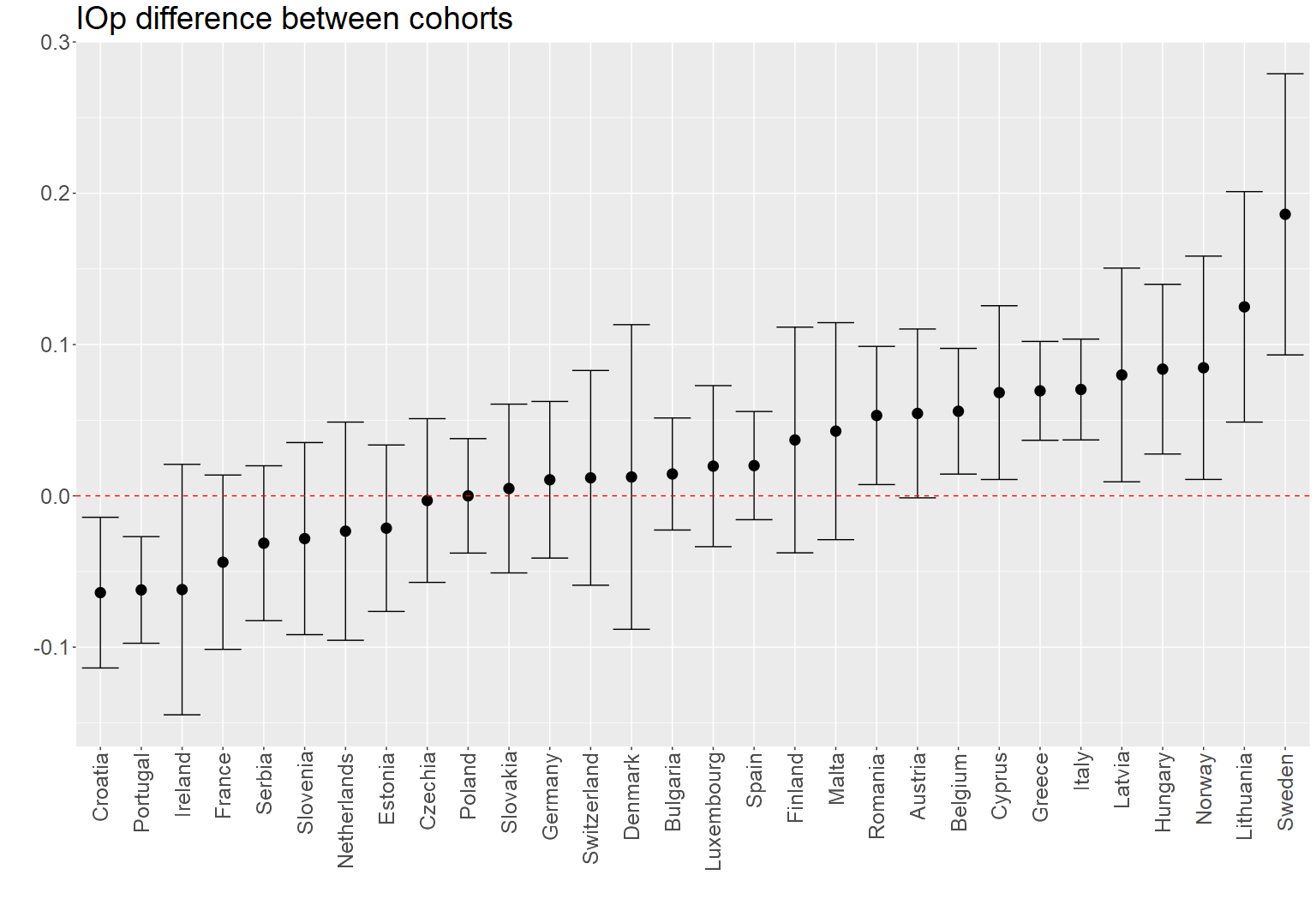}
    \caption{\small Difference between 1959-1975 and 1976-1994}
    \label{fig_cohortdiff}
\end{figure}

The sample includes individuals born from 1959 up to 1994. To compare more similar cohorts we divide the sample in two and estimate IOp in each cohort. The old cohort is formed by individuals born from 1959-1979 and the young cohort is born from 1976-1994. In both cohorts relative IOp still ranges from 30\% to 70\%. As we can see in Figure \ref{fig_cohortdiff}, relative IOp has not significantly varied from one cohort to the other for many countries. However, we see it has decreased for Croatia and Portugal and it has significantly increased for 10 countries; specially in Sweden where relative IOp in the young cohort is almost 20 percentage points higher.

Changes in IOp across cohorts can be due to several factors. First, there can be a composition effect. Increasing educational levels across cohorts means that the composition of circumstances change. Also, the returns to the same circumstances can change, for instance, the premium for having educated parents might increase or decrease across cohorts. Finally, unobserved circumstances which correlate with the observed ones can also change both in composition and returns. For instance, we do not observe parental income. If the correlation of our circumstances with income changes, the predictive power of our circumstances might also change to the extent we get predictive power from this correlation.

\subsection{Partial effects}

Table \ref{table_rel_gini_partial_effects} shows the largest relative partial effects for each country. Mother's education has the largest relative partial effect in 17 out of the 30 countries. This means that in a majority of countries, IOp decreases the most when we exclude mother's education. For instance, in Sweden excluding mother's education decreases IOp a 13\%. The next most common circumstance which attains largest IOp relative partial effect is sex. Sex obtains largest partial effects in 7 of the 30 countries. It seems to be the most important circumstance in all Baltic countries (Estonia, Latvia and Lithuania) and in several of the Nordic countries (Denmark, Finland and Norway). In Finland for instance, IOp decreases almost by a 19\% if we exclude sex. This seems to be driven by the superior academic attainment of females compared to males in these countries. Father's education has the biggest impact in Italy and father's occupation has the largest effect in Netherlands and Romania. Finally, tenancy has the largest partial effect in Germany and Malta and the number of adults working in the household when the individual was 14 has the largest partial effect in Portugal. 

In sum, in almost two thirds of the countries, parental education seems to be the most important predictor for education in terms of IOp. Hence, it seems that mobility plays a very important role. The interested reader can see partial and relative partial effects for all countries and circumstances in the Supplementary material.

\begin{table}[h!]
\centering
\footnotesize
\begin{tabular}{@{}cccccc@{}}
\toprule
Country & Max. PE & Max. Circumstance & Country & Max. PE & Max. Circumstance \\ \midrule
Austria & 0.044 & Mother's education & Ireland & 0.084 & Mother's education \\
 & (0.043, 0.045) &  &  & (0.083, 0.086) &  \\
Belgium & 0.027 & Mother's education & Italy & 0.029 & Father's education \\
 & (0.026, 0.028) &  &  & (0.028, 0.029) &  \\
Bulgaria & 0.018 & Sex & Lithuania & 0.072 & Sex \\
 & (0.017, 0.019) &  &  & (0.07, 0.073) &  \\
Switzerland & 0.072 & Mother's education & Luxembourg & 0.024 & Mother's education \\
 & (0.07, 0.073) &  &  & (0.023, 0.025) &  \\
Cyprus & 0.028 & Mother's education & Latvia & 0.169 & Sex \\
 & (0.027, 0.029) &  &  & (0.167, 0.171) &  \\
Czechia & 0.097 & Mother's education & Malta & 0.052 & Tenancy \\
 & (0.096, 0.098) &  &  & (0.05, 0.053) &  \\
Germany & 0.061 & Tenancy & Netherlands & 0.036 & Father's occupation \\
 & (0.06, 0.062) &  &  & (0.035, 0.037) &  \\
Denmark & 0.118 & Sex & Norway & 0.048 & Sex \\
 & (0.116, 0.12) &  &  & (0.046, 0.051) &  \\
Estonia & 0.157 & Sex & Poland & 0.05 & Mother's education \\
 & (0.156, 0.159) &  &  & (0.05, 0.051) &  \\
Greece & 0.018 & Mother's education & Portugal & 0.037 & Number of adults working \\
 & (0.018, 0.019) &  &  & (0.036, 0.038) &  \\
Spain & 0.033 & Mother's education & Romania & 0.028 & Father's occupation \\
 & (0.032, 0.033) &  &  & (0.027, 0.028) &  \\
Finland & 0.186 & Sex & Serbia & 0.043 & Mother's education \\
 & (0.184, 0.188) &  &  & (0.042, 0.043) &  \\
France & 0.032 & Mother's education & Sweden & 0.133 & Mother's education \\
 & (0.031, 0.033) &  &  & (0.131, 0.136) &  \\
Croatia & 0.056 & Mother's education & Slovenia & 0.089 & Mother's education \\
 & (0.056, 0.057) &  &  & (0.088, 0.091) &  \\
Hungary & 0.072 & Mother's education & Slovakia & 0.065 & Mother's education \\
 & (0.071, 0.073) &  &  & (0.064, 0.066) &  \\ \bottomrule
\end{tabular}
\caption{\small Largest Gini IOp relative partial effects.}
\label{table_rel_gini_partial_effects}
\end{table}

\subsection{Mobility: the role of mother's education}

In line with the results from the partial effects we turn now to study intergenerational mobility and its relationship with IOp. To study mobility we will look into the relationship between mother's education and individual's educational attainment. Mother's education comes only in three levels in EUSILC: low (less than primary, primary education or lower secondary education), medium (upper secondary education and post-secondary non-tertiary education) and high (short-cycle tertiary education, bachelor's or equivalent level, master's or equivalent level, doctoral or equivalent level). As a measure of intergenerational persistence we regress years of education on this three level variable and take the slope coefficient.

\begin{figure}[!htb]
    \centering\includegraphics[scale = 0.35]{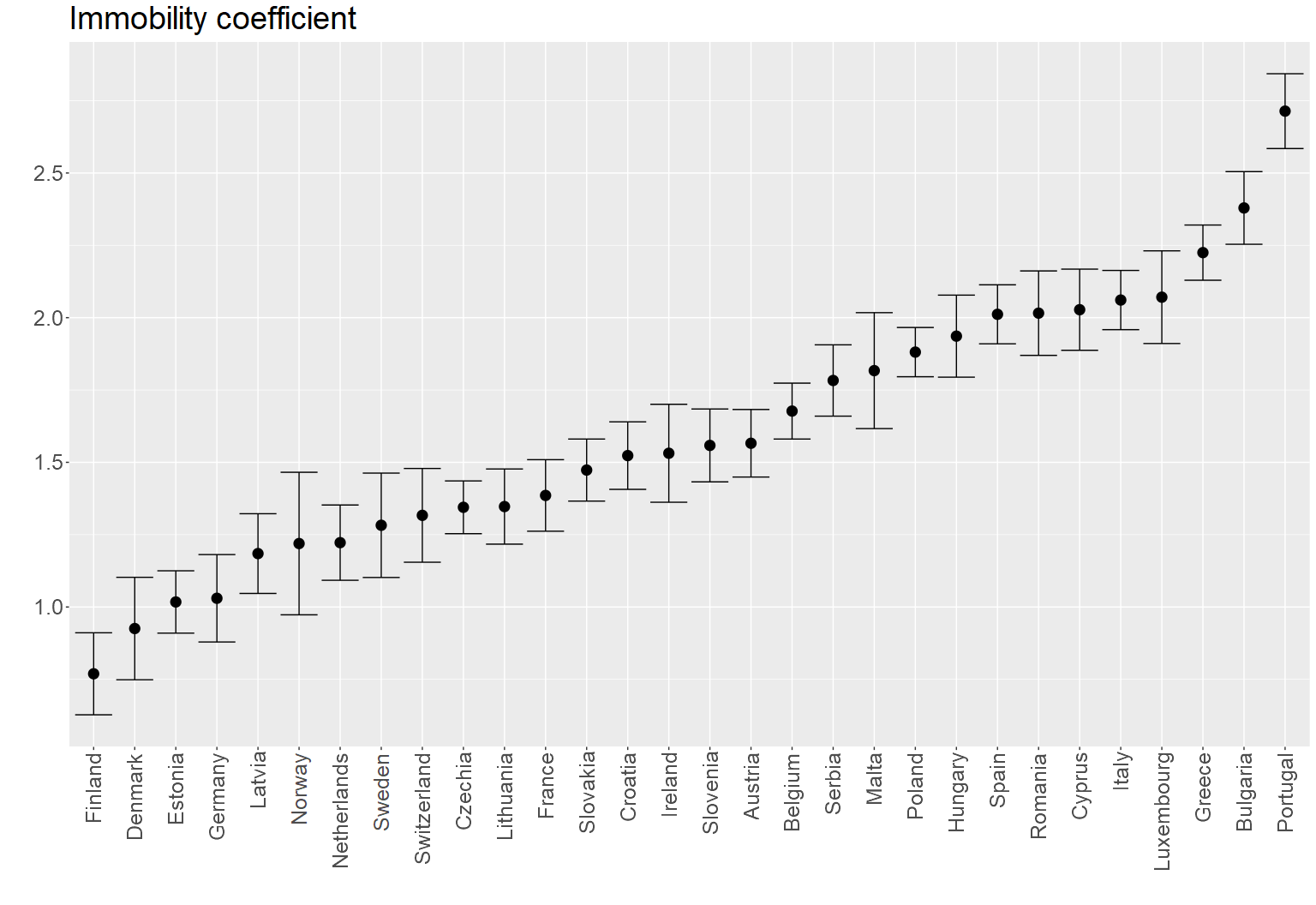}
    \caption{\small Slope coefficient of regressing years of education on mother's education}
    \label{fig_mobility}
\end{figure}

In Figure \ref{fig_mobility} we plot this measure of intergenerational persistence for all countries. A high value means there is low intergenerational mobility. We see that there is substantive intergenerational transmission of educational attainment. An extra level of mother's education is associated with almost a year more of educational attainment in Finland or more than $2.5$ years more of education in Portugal. Once more, we have Nordic countries with less persistent intergenerational educational attainment and we see southern countries, among others, having the greatest intergenerational dependence. 

In line with \cite{blanden2013cross}, \cite{corak2013income}, \cite{oecd2018} or \cite{BLANDEN2023405}, we look into the interrelationship between economic inequality and educational mobility. Economic inequality might cause parents with more resources to spend more on their childrens education compared to other parents. In turn, lower intergenerational mobility follows and we get an educational Great Gatsby curve. We take the income inequality results in \cite{escanciano2022debiased} to construct such educational Great Gatsby curves.

\begin{figure}[!htb]
    \centering\includegraphics[scale = 0.35]{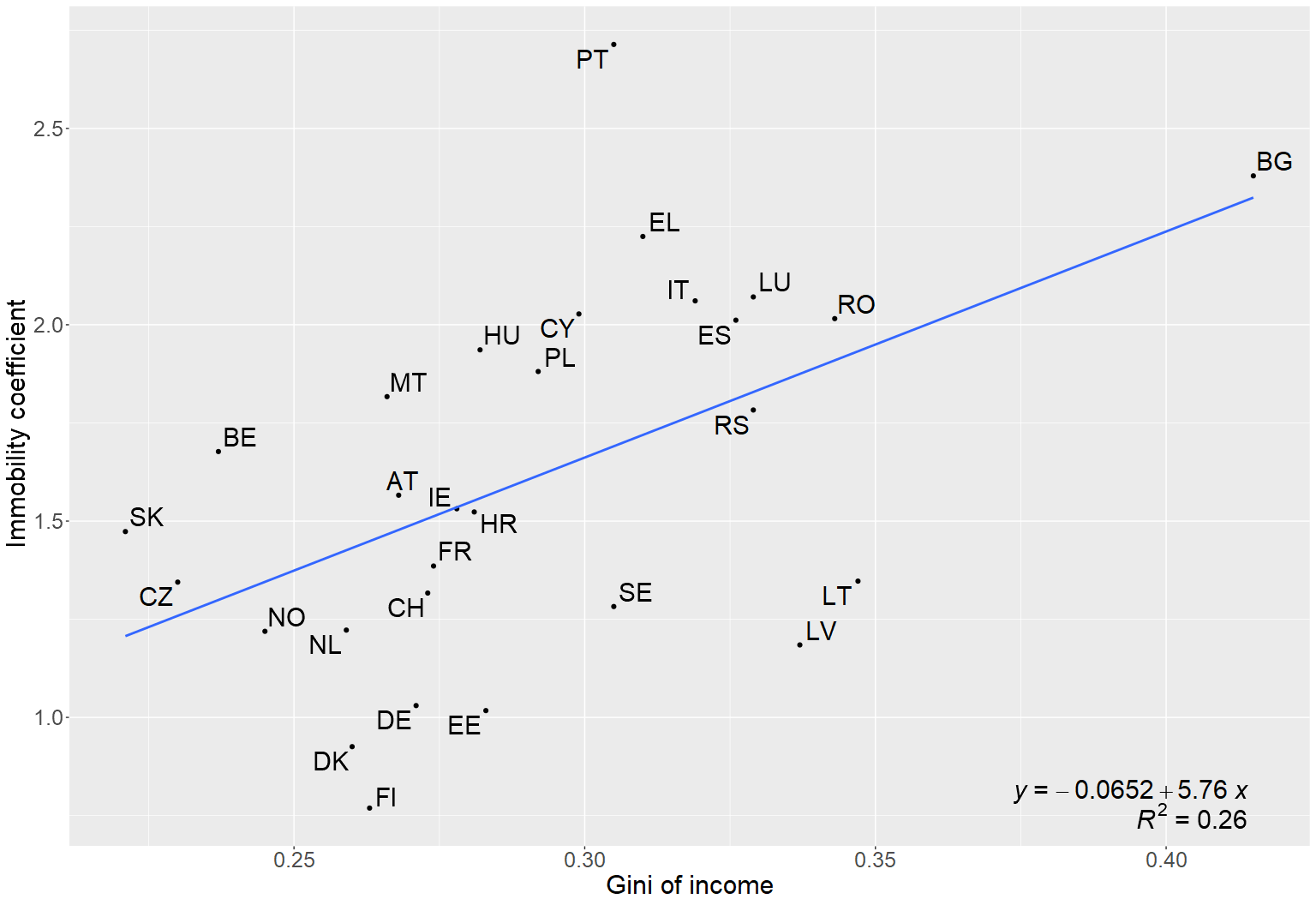}
    \caption{\small Education Great Gatsby Curve}
    \label{fig_GG_mobility}
\end{figure}

In Figure \ref{fig_GG_mobility} we see that there is a positive association between economic inequality and intergenerational transmission of educational attainment as reported in previous studies (see \cite{BLANDEN2023405}). This relationship is clearly endogenous. On the one hand, higher levels of economic inequality cause a greater disparity in financial constraints and parental educational investment capabilities. Hence, to the extent that the more educated parents are at the top of the income distribution, economic inequality will lead to less educational mobility. These mechanisms have been highlighted in economic models as in \cite{becker1979equilibrium} (see \cite{BLANDEN2023405} for more insights on modelling economic inequality and mobility). On the other hand, less educationally mobile societies will transmit the inequalities in the educational system to the labor markets.

Interestingly, the mechanism connecting educational immobility to economic inequality can only work through income IOp. In the absence of income IOp, individuals with highly educated parents should not have different income distributions. However, since we do know that there are returns to education, educational immobility means some individuals will earn more thanks to their parental education. This increase in economic inequality will be fully accounted by an increase in IOp. Conversely, IOp captures income inequalities which are explained by circumstances. If people with advantageous circumstances generally fare better, not only in income but also generally; we would expect countries with high income IOp to be less mobile. For instance, if the observed circumstances form a proxy for some latent status, then high income IOp would mean that high status families have more resources to devote to their offspring education. Hence, children in such families would have a double advantage, that of a higher status and that of higher monetary investments in their education. For an account of latent status and mobility see \cite{stuhler2023}. Hence, an association between income IOp and educational intergenerational mobility can also be established both ways. This motivates the construction of an IOp aware educational Great Gatsby curve.

\begin{figure}[!htb]
    \centering\includegraphics[scale = 0.35]{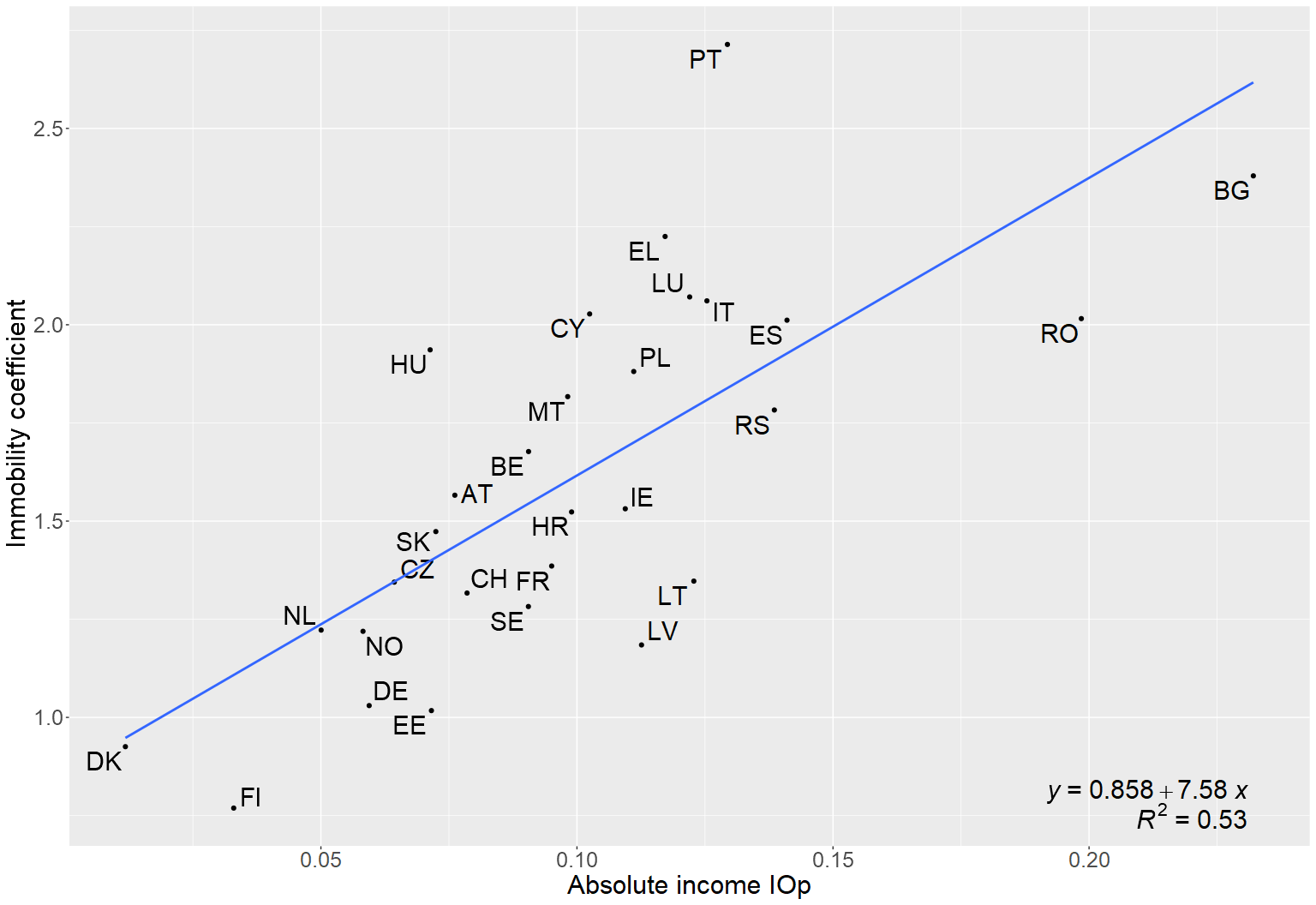}
    \caption{\small IOp aware Education Great Gatsby Curve}
    \label{fig_GG_opp_mobility}
\end{figure}

In Figure \ref{fig_GG_opp_mobility} we see such curve. To make the units comparable with the previous figure we have absolute income IOp in the $x$ axis, i.e. the Gini of the income predictions and not the share of income inequality explained by circumstances. We see that an increase of $0.01$ (a Gini point) in absolute income IOp is associated with an increase in the immobility coefficient of almost $0.08$ years (almost a month) while an increase of one Gini point in income inequality was associated to an increase in the immobility coefficient of almost $0.06$ years (three quarters of a month). Also, the $R^2$ doubles. Some correlation is expected to occur mechanically since IOp depends on own and mother's education as well as on many other circumstances which are correlated with educational variables. However, this mechanical correlation can only happen to the extent that IOp changes with the distribution of mother's education. 

Finally, we consider the relationship between educational IOp and income IOp. Again, both measures are determined by the same variables and hence correlation is expected. However, this is so to the extent that both measures are monotonic in the same circumstances. If there are countries where certain circumstances are related to high income IOp and high education IOp, then we will see this relation. One case in which one could expect there to be no correlation is if there were countries with educational institutions which manage to provide the same quality of education to all regardless of their background, where there are no financial barriers to higher education and where aspirations do not change across family backgrounds. These countries could have low educational IOp but still have high income IOp if with the same educational attainment, workers from more advantageous families get better results in the labor market.

\begin{figure}[!htb]
    \centering\includegraphics[scale = 0.35]{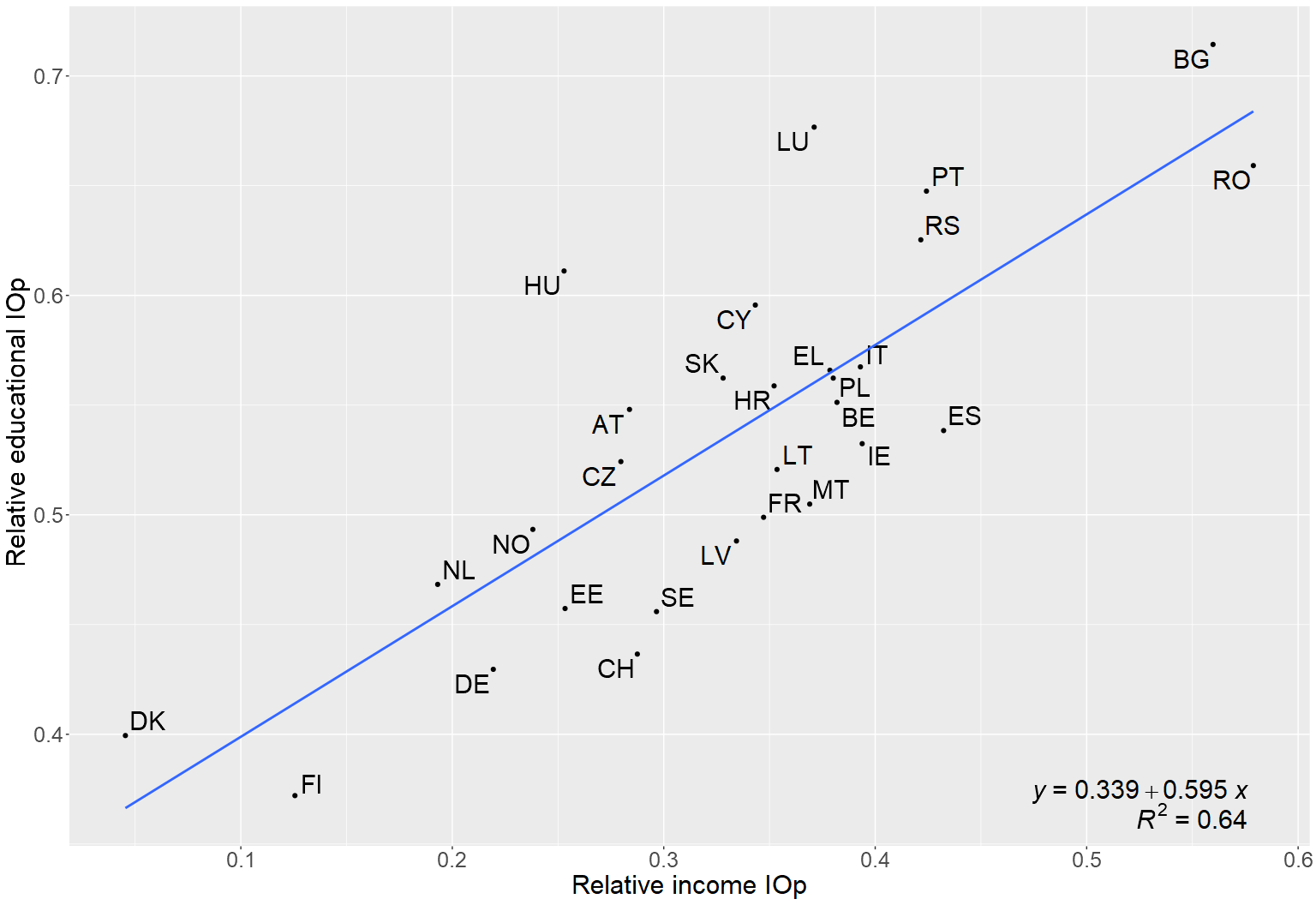}
    \caption{\small Educational IOp vs income IOp}
    \label{fig_iop_educ_iop_inc}
\end{figure}

In Figure \ref{fig_iop_educ_iop_inc} we see that countries where income IOp is high also have high levels of educational IOp. Asserting causality is extremely complicated, however, the correlation suggests that the observed circumstances do form part of a status that permeates income and educational inequalities making both income and educational attainment predictable by the same circumstances. Also, there are no countries who manage to have low levels of educational IOp despite high levels of income IOp. This might be indicative that financial barriers matter and that educational quality might also change with family income and across family backgrounds. This would mean that educational institutions could have a hard time decreasing educational IOp without policies which also reduce income IOp.

\subsection{Gender educational IOp}

\begin{figure}[!htb]
    \centering\includegraphics[scale = 0.35]{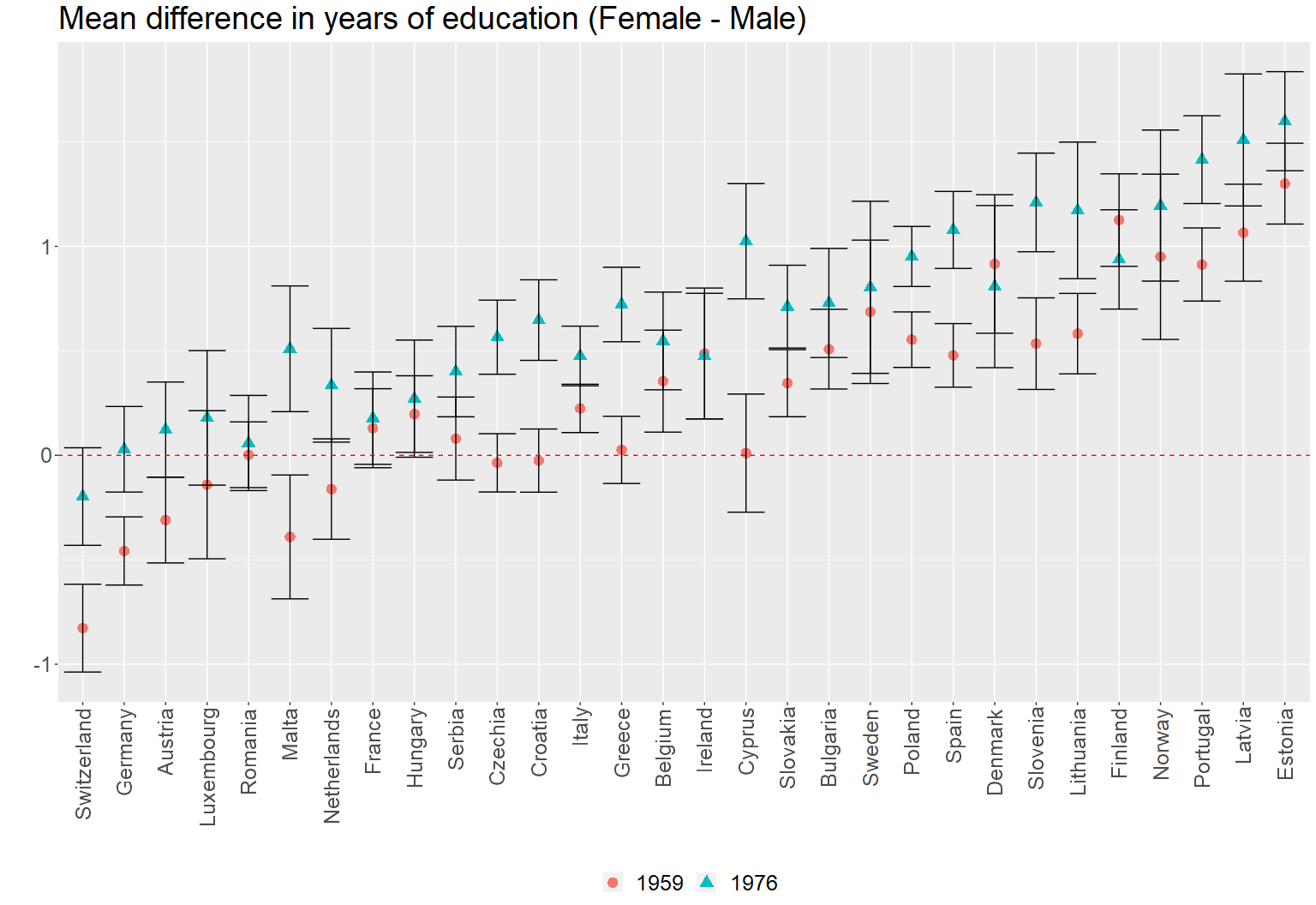}
    \caption{\small Average years of education by gender}
    \label{fig_educ_sex_mean}
\end{figure}

Now we analyse with more detail the gender aspect of educational IOp. Let us first take a look at mean differences in educational achievement for the young and the old cohort across sexes in Figure \ref{fig_educ_sex_mean}. We see that generally the difference between female average years of education and male average years of education is smaller for old cohorts. In fact, for almost half of the countries this difference is negative or not statistically different from zero for the old cohort. For the rest of the countries females acquired more years of education on average even in the old cohort. For the young cohort we see that in most countries females studied more years. This is specially so in the countries where sex had the largest relative IOp partial effect. In Estonia, Latvia, Lithuania, Denmark, Finland and Norway; females averaged one year or more of education.

Now we turn to differences in IOp. We want to see whether inequality among males can be explained by circumstances to a greater extent that that of females, i.e. we focus on how circumstances explain inequality differently depending on sex. A good approach to do this would be to use counterfactuals. That is, to see what would be the IOp if all were female/male (ruling out general equilibrium effects), or what would be IOp among males if they were female. Adding and subtracting these counterfactuals one could construct the famous Kitagawa-Oaxaca-Blinder (KOB) decomposition (\cite{kitagawa1955components}, \cite{blinder1973wage} and \cite{oaxaca1973male}).

We argue that looking at the differences between the share of inequality explained by circumstances in the sample of females and in the sample of males (henceforth the IOp gender gap) is enough. The IOp gender gap is determined by differences in the distribution of circumstances in each group (composition term) and differences in the conditional expectation functions of each group (structural term). In our case the composition term is necessarily zero since gender is random and the circumstances are predetermined. For example, parental education has the same distribution for males and females. This is corroborated by the data, where the balance of the circumstances is nearly perfect (see Supplementary material).

Hence, the IOp gender gap must be driven by differences in the conditional expectation functions and not by different circumstance distributions. Such differences in the conditional expectation function could be caused by different returns to circumstances and different dependence of observed circumstances with unobserved ones. For instance, if the maternal/paternal role models impact differently girls than boys, then the returns to parental education, occupation, etc. can be markedly different for females and males in an adult age. Another example would be if, due to social rules, most males would generally pursue long STEM careers while females with parents related to STEM careers were much more likely to pursue a long STEM career than other females. This would create a heterogeneity in the career paths of females which would not be present in that of males; hence leading to a higher IOp among females.

\begin{figure}[!htb]
\centering
   \begin{minipage}{0.48\textwidth}
     \centering
     \includegraphics[width=1.1\linewidth]{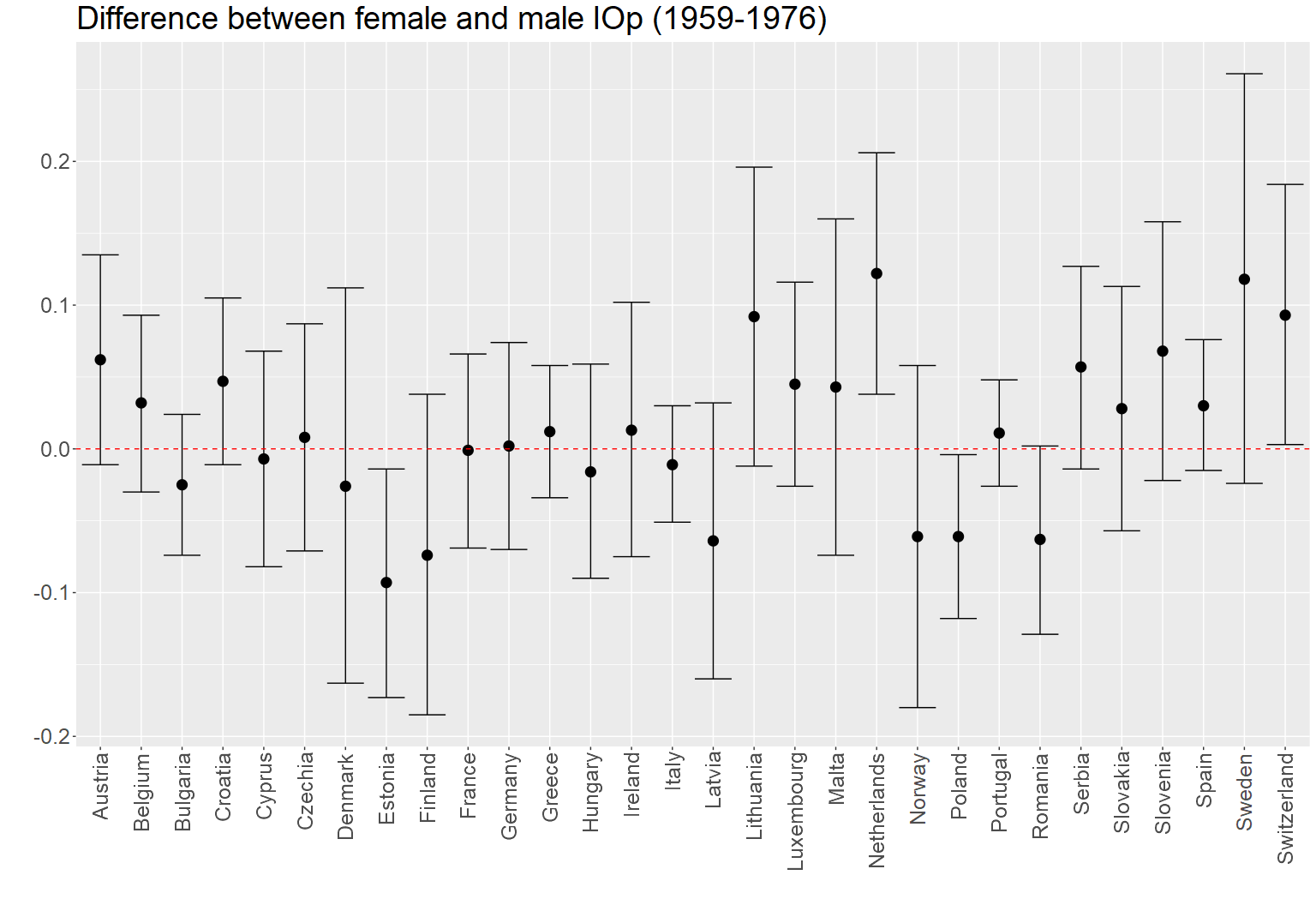}
     \caption{\small IOp Gender gap 1959-1976}\label{fig_IOpgap_old}
   \end{minipage}\hfill
   \begin{minipage}{0.48\textwidth}
     \centering
     \includegraphics[width=1.1\linewidth]{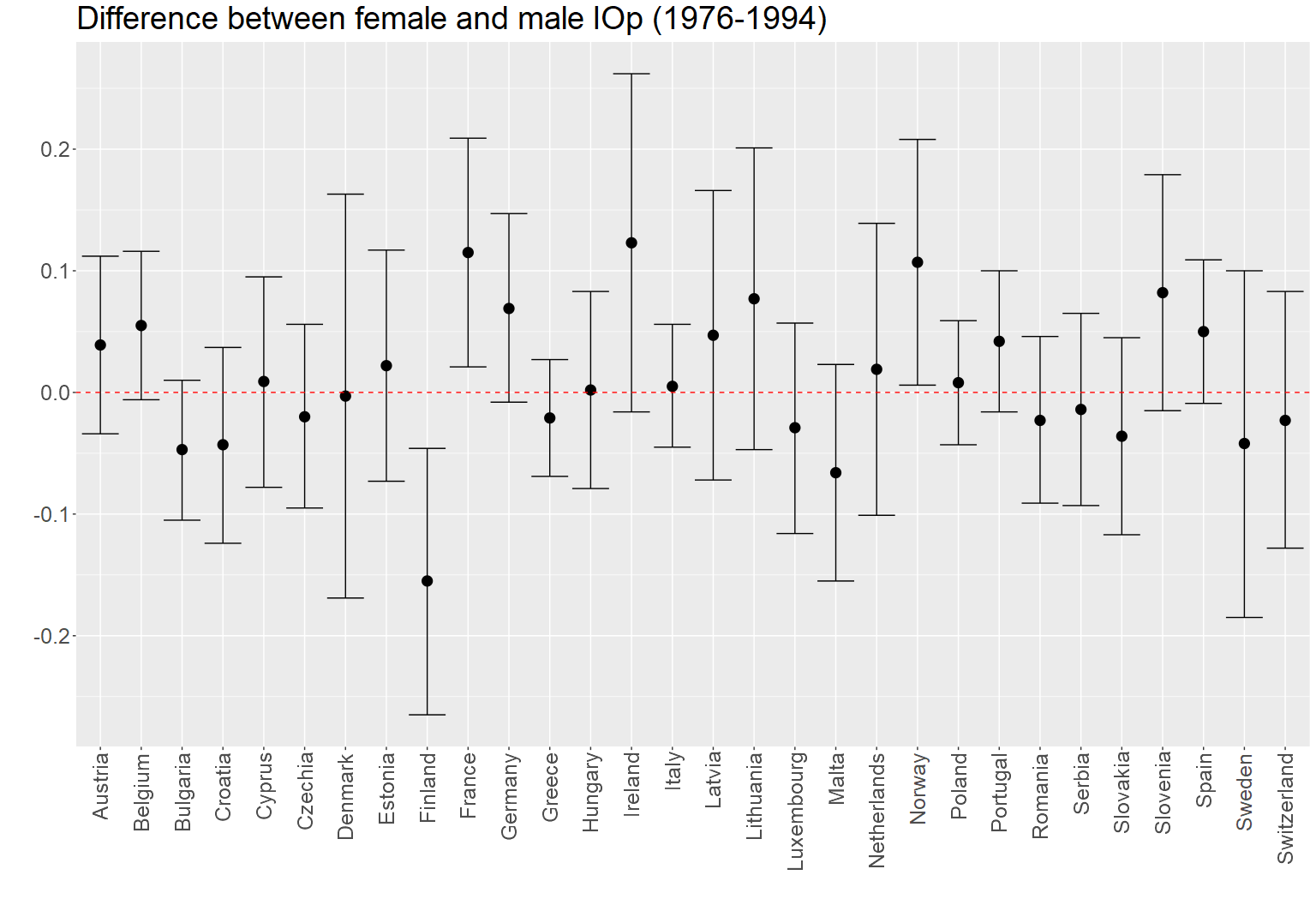}
     \caption{\small IOp Gender gap 1976-1994}\label{fig_IOpgap_young}
   \end{minipage}
\end{figure}

In Figures \ref{fig_IOpgap_old} and \ref{fig_IOpgap_young} we see the IOp gender gap for the old and the young cohort. For most countries we get that the gap is not significantly different from zero. This means that in most countries inequality within a sex is not explained by circumstances more than in the other sex. For the old cohort, we have that Estonia and Poland have a significantly negative IOp gap, meaning that IOp is higher among males than females. In Estonia, the males from the old cohort are more unequal than females, hence males have both higher inequality and higher IOp. In Poland males have greater IOp but lower inequality. In the old cohort, Netherlands and Switzerland have significantly positive IOp gaps meaning that IOp is higher among females. In the old cohort, inequality in the Netherlands is not significantly different but females have more IOp. In Switzerland females from the old cohort have both more inequality and higher IOp. In the younger cohort we see that only Finland has a negative IOp gap, males relative IOp is around 15 percentage points higher compared to female relative IOp. Young Finns have both higher inequality and higher IOp. For the young cohort the IOp gap is significantly positive for Norway (where inequality is higher among males) and for France (where inequality is not statistically different across sexes).

In general we see that there does not seem to be much of an IOp gap of any sign. In the countries where we do find one (Estonia, Poland, Netherlands and Switzerland in the old cohort and Finland, Norway and France in the young cohort) there does not seem to be any systematic correlation with the degree of inequality among each sex and the IOp gap.

Exploring the reasons why some countries have a negative, positive or zero IOp gap is an interesting question for further research which, unfortunately, is out of the scope of this paper. Also, it has to be kept in mind that the returns to circumstances might still differ even if the difference in IOp is zero. This is because some circumstances might predict more inequality for males and other circumstances for females and these might compensate each other. Further, we focus on years of education, still, the choice of educational field can differ markedly by gender. These qualitative differences in the field of study matter since different fields might have different returns. Finally, it is important to stress that we do not claim any kind of causality. It might be that our circumstances only predict well years of education because they are correlated with omitted variables which are the true causes. However, we claim this is not a problem for quantification of IOp since any such omitted variable will also be a circumstance.

\subsection{Further associations}

As a last exercise we see the correlations of our estimates with other interesting variables. Here we do not seek to find any causal relationships, we just want to highlight interesting associations in the data.

\begin{figure}[!htb]
    \centering\includegraphics[scale = 0.35]{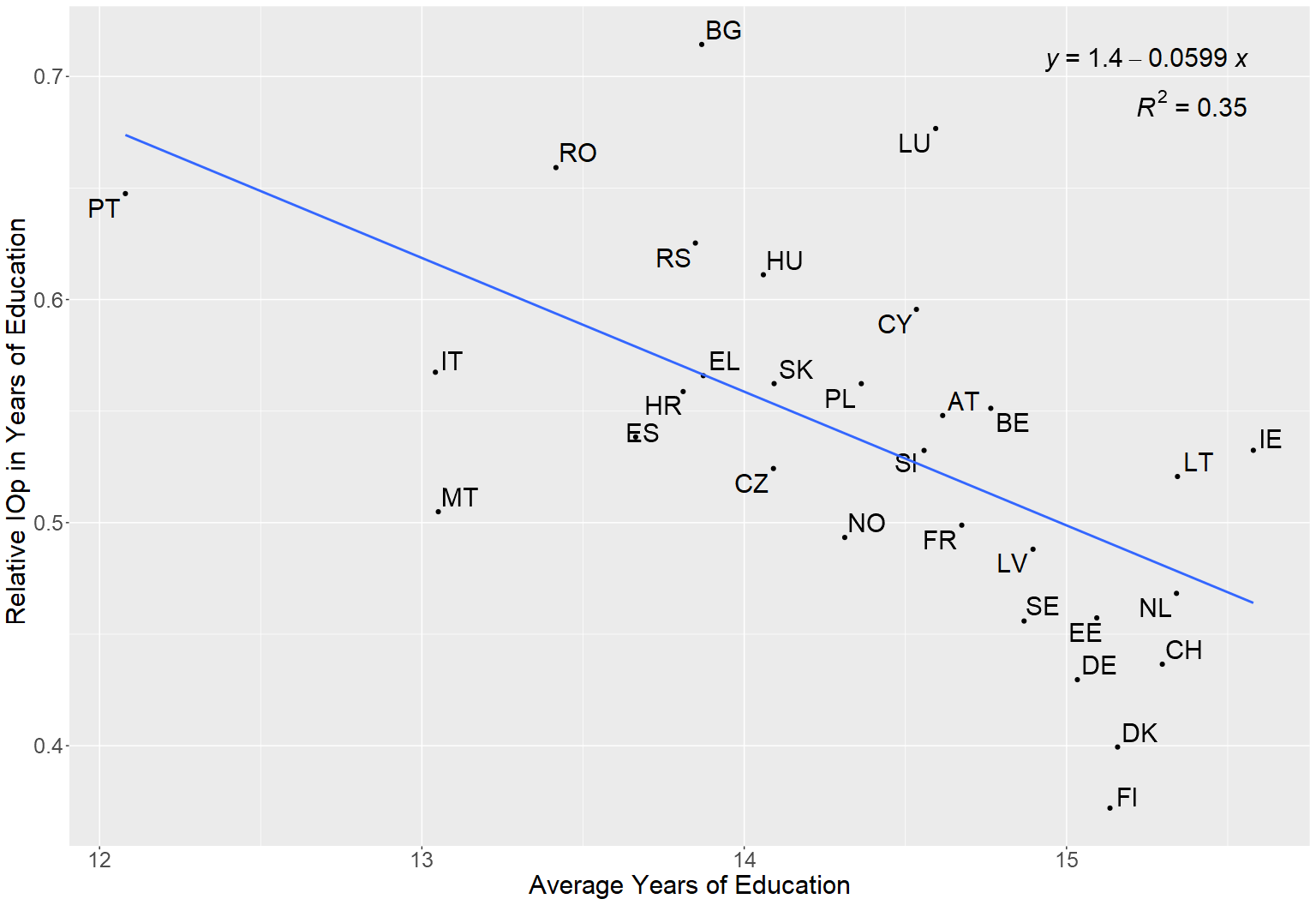}
    \caption{\small Educational IOp vs average years of education}
    \label{fig_iop_mean}
\end{figure}

\begin{figure}[!htb]
    \centering\includegraphics[scale = 0.35]{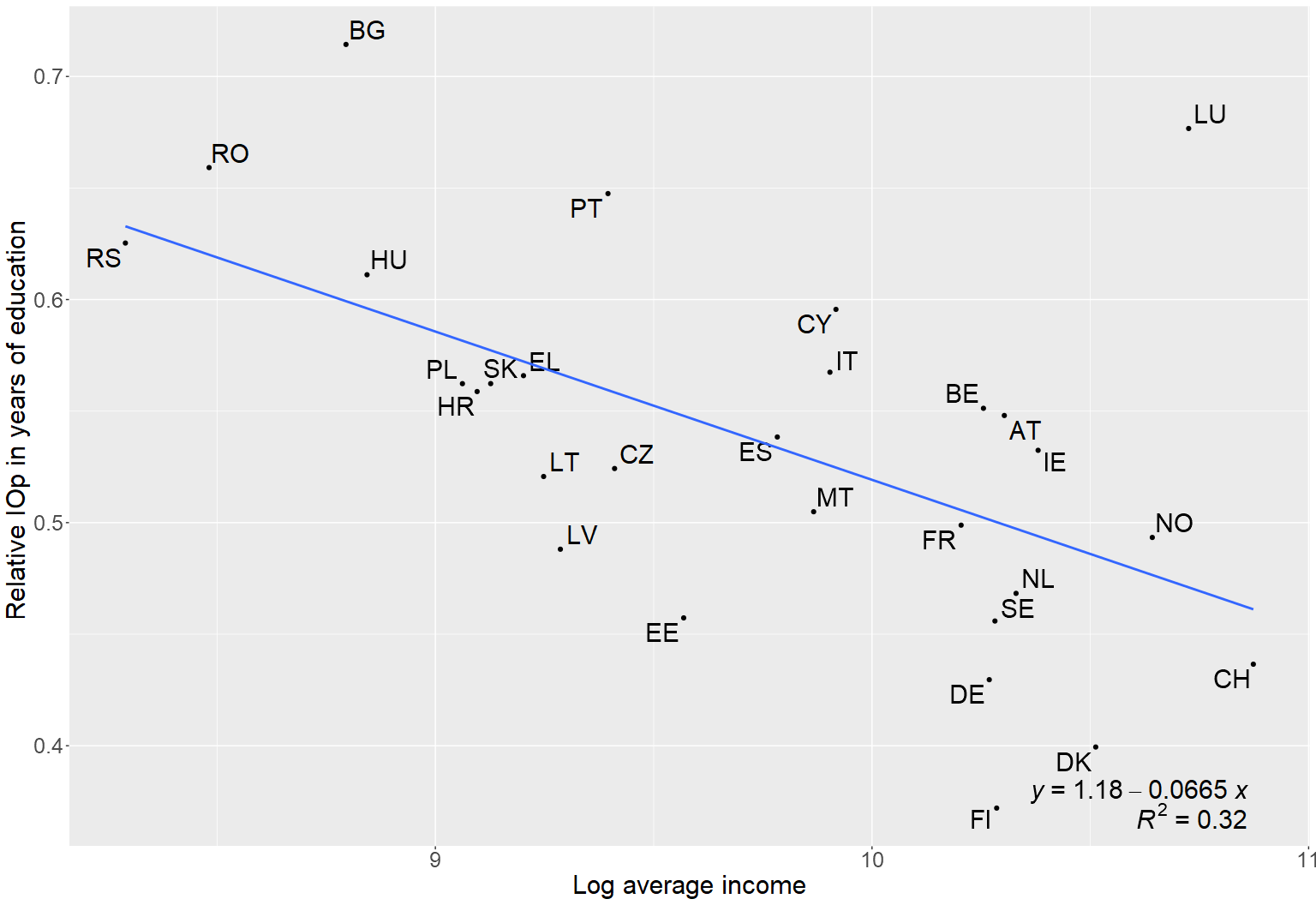}
    \caption{\small Educational IOp vs log average income}
    \label{fig_iop_meaninc}
\end{figure}

\begin{figure}[!htb]
    \centering\includegraphics[scale = 0.35]{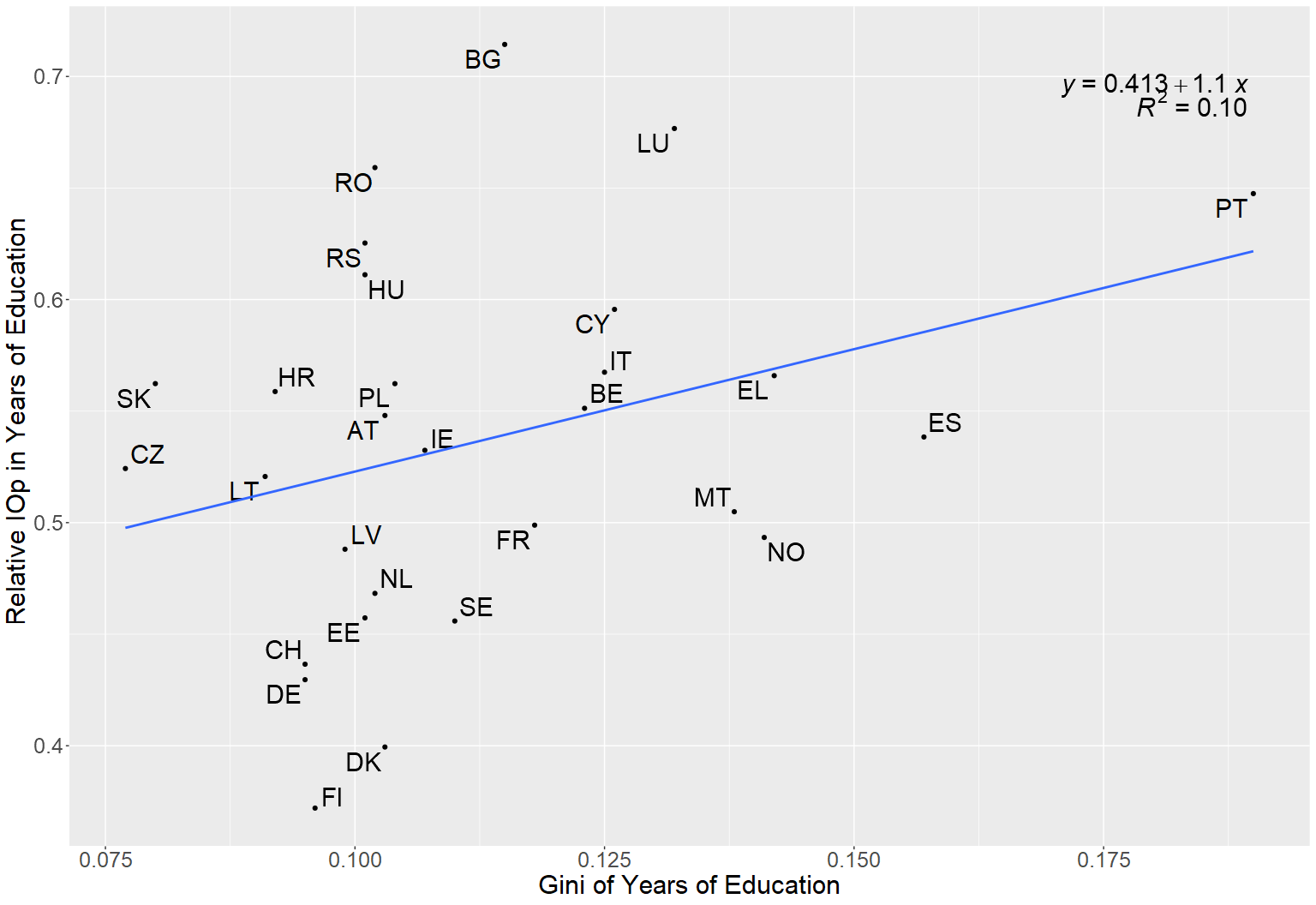}
    \caption{\small Educational IOp vs Gini of years of education}
    \label{fig_iop_gini}
\end{figure}

\begin{figure}[!htb]
    \centering\includegraphics[scale = 0.35]{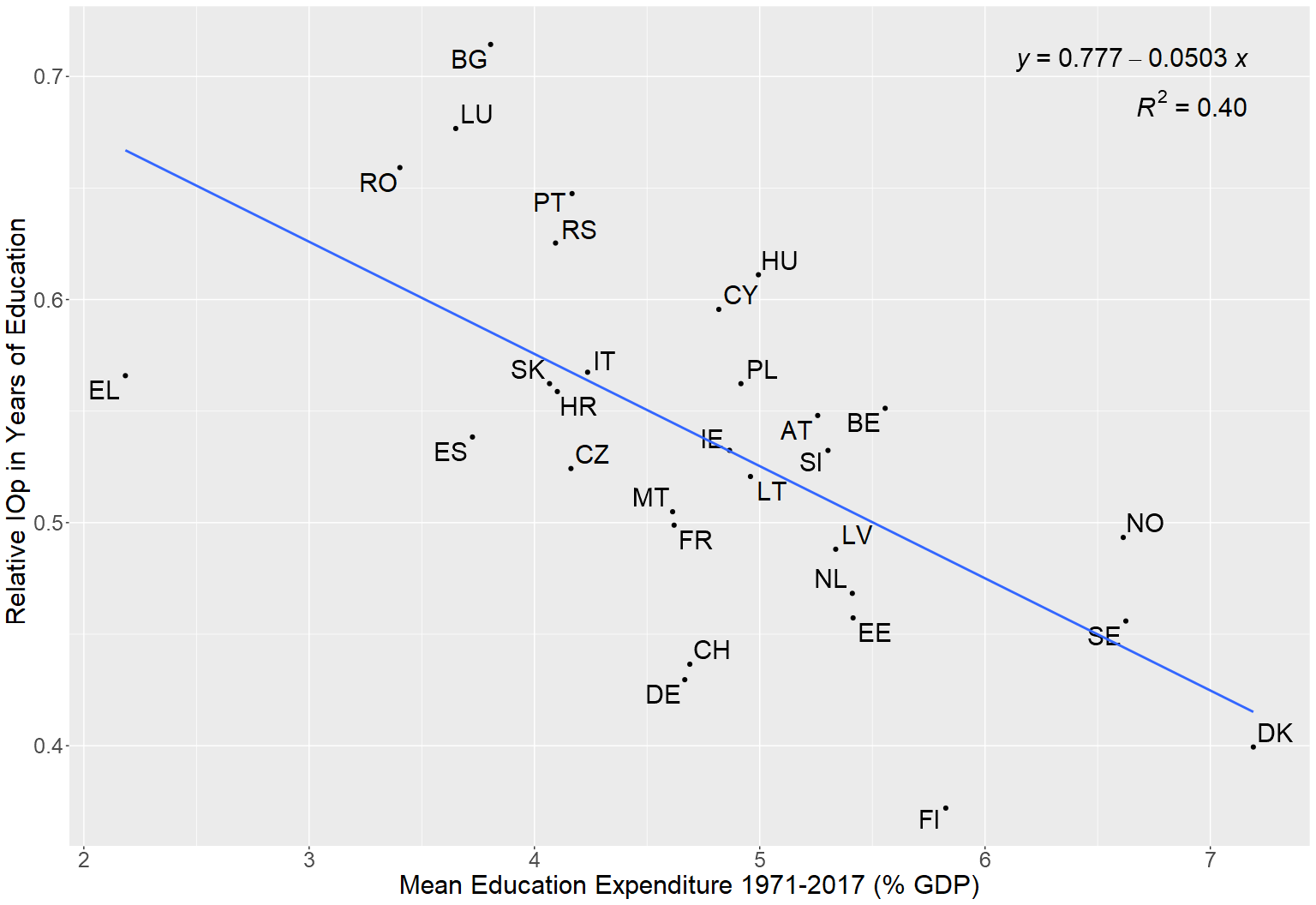}
    \caption{\small Educational IOp vs Mean education expenditure (\% GDP)}
    \label{fig_iop_educexp}
\end{figure}

In Figure \ref{fig_iop_mean}, we see that countries where average years of education is high tend to have lower levels of IOp. This could be driven by the fact that we are top coding years of education in 18 years. However, to the extent that we can consider a reasonable maximum level of years of education, compressing the distribution towards this maximum can have the effect of making education less sensible to circumstances. In Figure \ref{fig_iop_meaninc} we see that a 1\% increase in average income is associated with almost 0.07 percentage points decrease in relative IOp. Hence, richer societies tend to have lower levels of educational IOp. In Figure \ref{fig_iop_gini} we look at the correlation between IOp and the Gini in years of education. We see a positive slope although the relationship is not as strong as in Figure \ref{fig_iop_mean}. Generally, the more unequal the distribution of years of education the more IOp there seems to be. Still, some countries such as Romania or Bulgaria have lower inequality which is explained to a great extent by circumstances or countries like Spain have high inequality but lower IOp than many other countries. In Figure \ref{fig_iop_educexp} we see a strong negative correlation between education expenditure as a percentage of GDP and IOp. 

\begin{figure}[!htb]
    \centering\includegraphics[scale = 0.35]{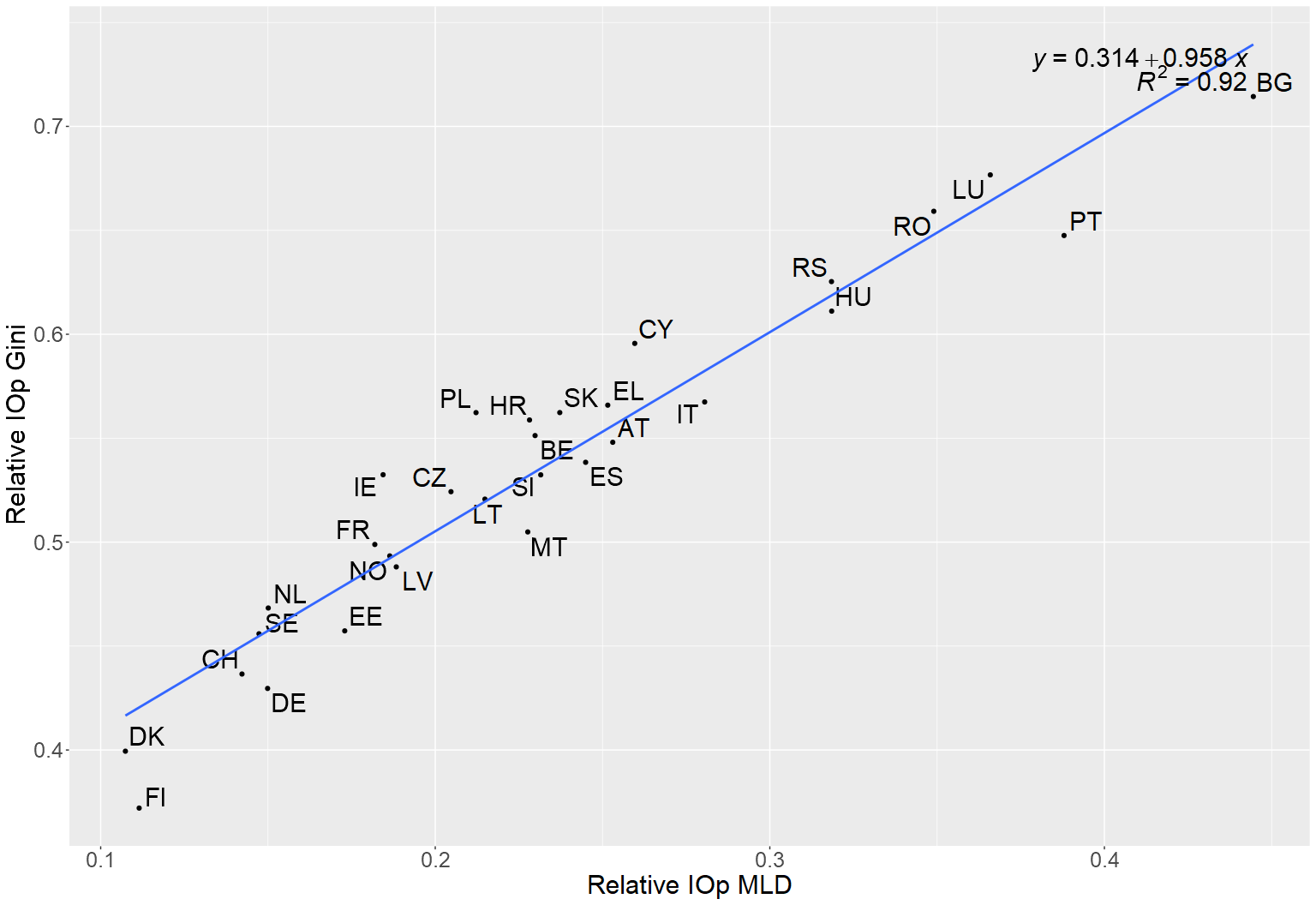}
    \caption{\small Relative IOp Gini vs Relative IOp MLD}
    \label{fig_sct_iopgini_iopmld}
\end{figure}

Finally, we see the correlation between relative IOp measured by the Gini and relative IOp measured by MLD in Figure \ref{fig_sct_iopgini_iopmld}. It is reassuring to see that the correlation is almost $1$ and the $R^2$ is 0.92. As expected, MLD estimates lower levels of relative IOp due to the fact that extreme predicted values are uncommon and the MLD cares mostly about inequalities in the tails. Nevertheless, up to a difference in levels, both measures seem to be capturing IOp in a similar manner. All the results in this paper can be found in the Supplementary Appendix for the MLD.

\section{Discussion}
\label{sec_discussion} 

It is important not to attribute a causal interpretation to these findings. For instance, the circumstance attaining the highest relative partial effect in Portugal is the number of adults working in the household when the individual was 14 years old. If there is any unobserved circumstance which is correlated with this circumstance and actually causes the disparity in incomes, it could be that the number of adults is just reflecting the predictive power of that unobserved circumstance.

Also, we focus on educational attainment (years of education) but this leaves out many important dimensions. If we look at educational achievement (i.e. grades) the results could change considerably. For instance, \cite{BLANDEN2023405} do not find any clear relationship between economic inequality and educational achievement while they do find it for educational attainment. This happens for instance if two students perform equally well in high school but the most advantaged student goes on to university and the most disadvantaged does not, for example due to financial barriers.

Another dimension of education which this study does not take into account is the heterogeneity in quality of the educational institutions and the differential choice of fields of study. \cite{chetty2017mobility} find out that those in the top of the income distribution are much more likely to attend higher quality education institutions such as an Ivy League college. \cite{hallsten2018multiple} find that a quarter of the variation in the choice of field study can be explained by parental background conditional on the same previous achievement. \cite{kim2015field} finds that the choice of educational field later matters in terms of returns. For a more comprehensive review see \cite{BLANDEN2023405}. Hence, in this study we are treating individuals with the same years of education the same when they are in many occasions qualitatively different. Some of them might have attended the same amount of years to a top quality educational institution or studied a field which generates much higher returns. The inequalities stemming from these qualitative differences will be in turn shaped by the circumstances we observe. 

Furthermore, for computational reasons we have only tuned the penalization parameters in Lasso and Ridge. The tuning parameters of the other MLs are very close to the default ones. The parameters are tuned in the same way for all countries. Exploration into better parameter tuning could increase the precision of predictions in the first step. Nevertheless, local robustness makes debiased IOp measures less sensitive to tuning parameters.

Finally, one could have considered ensemble methods where all MLs are combined into one giving weights to each ML in a cross-validated optimal way. This can be done with the R package \texttt{SuperLearner}. The R package provided in this paper provides this option but it is computationally expensive so we did not make use of it for this empirical application.

\section{Conclusion}
\label{sec_conclusion}
We find that circumstances can explain a large share of educational attainment inequality in European countries. In most countries, mother's education is the most important variable to explain educational inequality. Motivated by this fact, we study the intergenerational mobility in educational attainment and we confirm that there is substantive intergenerational immobility in educational attainment. Further, countries with high income inequality display less intergenerational mobility. This association is strengthened when we replace income inequality by income IOp. When we relate educational attainment IOp and income IOp we also find a strong relationship. It is hard to find countries who have managed to have a low educational IOp and a high income IOp. This points to the fact that these two problems need to be addressed jointly.  

\newpage

\bibliographystyle{ectabib}
\bibliography{references}

\end{document}